\documentclass[ 
reprint, 
superscriptaddress,
amsmath,
amssymb, 
floatfix,
longbibliography,
aps,
prl,
]{revtex4-1}

\usepackage{graphicx}
\usepackage{xcolor}
\usepackage[colorlinks=true,citecolor=blue,linkcolor=magenta]{hyperref}
\usepackage{bm}

\usepackage{mathtools}
\newtheorem{theorem}{Theorem}

\begin{document}

\title{Extracting randomness from magic quantum states}
\author{Christopher Vairogs}
\affiliation{Theoretical Division, Los Alamos National Laboratory, Los Alamos, New Mexico 87545, USA}
\affiliation{Department of Physics, University of Illinois at Urbana-Champaign, Urbana, Illinois 61801, USA}
\author{Bin Yan}
\affiliation{Theoretical Division, Los Alamos National Laboratory, Los Alamos, New Mexico 87545, USA}

\date{\today}

\begin{abstract}

Magic quantum states (non-stabilizer states) play a pivotal role in fault-tolerant quantum computation. Simultaneously, random resources have emerged as a key element in various randomized techniques within contemporary quantum science. In this study, we establish a direct connection between these two notions. More specifically, our research demonstrates that when a subsystem of a quantum state is measured, the resultant unmeasured part of the system can exhibit a high degree of randomness that can be enhanced by the inherent correlations of the underlying magic quantum state. 
Our findings suggest an approach to quantifying correlations within magic quantum states beyond the conventional paradigm of entanglement, and introduce an efficient approach for leveraging such correlations to generate random quantum resources.
\end{abstract}

\maketitle

The field of quantum computing has experienced remarkable advancements in recent decades~\cite{Arute2019,Zhong2020,Bluvstein2023,Preskill2018}, yet the essential source of power driving the quantum speedup continues to be elusive. This complex issue is inherently multifaceted, with its resolution dependent on the theoretic framework for characterizing the underlying resources. For instance, entanglement was known to be insufficient --- Clifford circuits produce entanglement among qubits but nevertheless can be efficiently simulated classically~\cite{Gottesman1998}. On the flip side, the same set of operations constitutes the basic building block for the stabilizer formalism of quantum computing, which, when augmented with the so-called {\it magic quantum states}~\cite{Bravyi2005,Knill2005,Campbell2012,Howard2014}, makes a blueprint for universal fault tolerant quantum computation. This paradigm thus recognizes magic states as a crucial quantum resource. The exploration and exploitation of the physical properties of quantum magic states have recently emerged as an active area of research\textcolor{red}{~\cite{Veitch2014, Howard2017, Wang2020, Leone2022}}. 

Randomness stands as another pivotal resource in quantum information science, playing a crucial role in numerous protocols for quantum information processing. Various applications, including quantum device benchmarking~\cite{Arute2019,Cross2019,Neill2018,Harris2022,Wu2021}, tomography~\cite{Huang2020,Elben2023} and quantum channel approximation~\cite{Hayden2004,Yan2022,Kunjummen2023}, rely on random quantum resources. In a recent breakthrough~\cite{Cotler2023,Choi2023}, it was demonstrated that random resources can be extracted from complex many-body quantum states. Specifically, when a subsystem of a strongly correlated many-body state undergoes projective measurements, the unmeasured subsystem yields a highly random quantum state ensemble, which nevertheless may not be directly prepared efficiently. This result not only offers an efficient protocol for generating random quantum states as a practical resource but also opens up new avenues for exploring fundamental physical problems, such as non-equilibrium quantum dynamics~\cite{Ippoliti2022,Lucas2023,Ippoliti2023}, and, as will be delved into in this work, characterization of quantum states.

\begin{figure}[b!]
    \centering
    \includegraphics[width=\columnwidth]{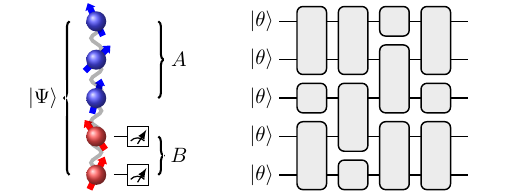}
    \caption{Left: A subsystem $B$ of a complex many-body state $|\Psi\rangle$ is subject to projective measurement. The unmeasured subsystem $A$ is projected to an ensemble of random states. Right: The qubits are prepared in a product magic state, parameterized by $\theta$, which is then tranformed by a random Clifford circuit, generating a state with the same magic value.}
    \label{fig:setup}
\end{figure}

In this context, we unveil a profound connection between randomness and magic (the degree of non-stabilizerness of a quantum state, as will be defined in the following). Specifically, we demonstrate that typical quantum states exhibiting higher magic values yield post-measurement projected ensembles with greater randomness. The degree of magic and randomness are precisely quantified using proper metrics---namely, stabilizer R\'{e}nyi entropy and quantum state $t$-designs. 

The significance of this result is two-fold: On a conceptual level, it unravels an intrinsic randomness associated with magic quantum states. This randomness is a manifestation of many-body correlations that are not captured by conventional measures such as entanglement. Consequently, these findings shed new light on the role of magic as a quantum resource. Simultaneously, it proposes an efficient scheme for leveraging such correlations for random quantum state generation with only Clifford circuit operations. The later comprises operations that are naively fault tolerant in quantum computing architectures protected by stabilizer error-correcting codes, and therefore operationally ``cost-free''. In the following, we will lay down the essential components of the setup and present the main theorem, as well as supporting evidence from numerical simulations.

\vspace{8pt}
\emph{Projected ensemble.}---
For any ensemble of pure quantum states, $\mathcal{E} = \{(p_i, |\psi_i\rangle)\}$, the level of its randomness can be quantified rigorously using the notion of quantum state design. Define the $t$-th moment of $\mathcal{E}$ as
\begin{equation}\label{eq:moment}
    \rho^{(t)}_{\mathcal{E}} = \sum_i p_i |\psi_i\rangle\langle \psi_i|^{\otimes t},
\end{equation}
the average of its $t$-th tensor power, and denote especially the $t$-th moment of the Haar random ensemble by $\rho_{\text{Haar}}^{(t)}$. We say that $\mathcal{E}$ is a \textit{quantum state $t$-design} if $\rho_{\mathcal{E}}^{(t)} = \rho_{\text{Haar}}^{(t)}$, that is, $\mathcal{E}$ is indistinguishable from Haar measure up to its $t$-th moment. In practice, an exact $t$-design can never be reached. It is therefore useful to have a measure of the approximate closeness of the ensemble's distribution to the Haar measure. This idea is captured by an \textit{$\varepsilon$-approximate quantum state $t$-design}, an ensemble $\mathcal{E}$ for which 
\begin{equation}\label{eq:distance-def}
    d^{(t)}(\mathcal{E}) \equiv ||\rho_{\mathcal{E}}^{(t)} - \rho_{\text{Haar}}^{(t)}|| \leq \varepsilon.
\end{equation}
Here, $||\cdot||$ can be any proper norm over density matrices. Throughout this work, we will employ two commonly used norms, namely, trace norm $||\rho||_1 \equiv \rm{tr}\sqrt{\rho^\dag\rho}$ and the Hilbert-Schmidt norm $||\rho||_2 \equiv \sqrt{\rm{tr}\rho^\dag\rho}$, and denote the corresponding distance as $d^{(t)}_{\rm trace}(\mathcal{E})$ and $d^{(t)}_{\rm HS}(\mathcal{E})$, respectively.

A basic ingredient of our work is the use of a subsystem measurement to generate an ensemble of random states (Fig.~\ref{fig:setup}, left). Consider a bipartite pure state $|\Psi\rangle\in\mathcal{H}_{AB} = \mathcal{H}_A \otimes \mathcal{H}_B$ over systems $A$ and $B$. We take the $A$ and $B$ subsytems to be composed of $N_A$ and $N_B$ qubits, respectively. Denote $N \equiv N_A + N_B$, $d_A \equiv 2^{N_A}$, $d_B \equiv 2^{N_B}$, and $d \equiv 2^N$, and let $\{|i\rangle_B\}_{i=1}^{d_B}$ be a basis of product states for $\mathcal{H}_B$. The outcomes of a projective measurement on the $B$ subsystem with respect to this basis, followed by a discarding of the $B$ subsystem, form an ensemble of states $\mathcal{E}_{|\Psi\rangle} \equiv \{(p_i, |\psi_i\rangle)\}$ on the $A$ subsystem, where the probability $p_i$ is
\begin{equation}
    p_i = \langle \Psi|(I_A \otimes |i\rangle\langle i|_B)|\Psi\rangle
\end{equation}
and the corresponding projected state of $A$ is
\begin{equation}
    |\psi_i\rangle = (I_A \otimes \langle i|_B)|\Psi\rangle/\sqrt{p_i}.
\end{equation}
We will refer to the ensemble $\mathcal{E}_{|\Psi\rangle}$ as the \textit{projected ensemble}~\cite{Anza2021} generated by $|\Psi\rangle$, whose statistics can be quantified by $t$-design discussed above.

Conceivably, the distribution of the projected ensemble reflects the intrinsic randomness of the underlying generating state; a more complex many-body state, e.g., with stronger entanglement, may produce a projected ensemble with a higher degree of randomness. Indeed, this intuition was captured formally in a recent result in~\cite{Cotler2023,Choi2023}, that is, if the pre-measurement state $|\Psi\rangle$ itself was sampled from an $\varepsilon'$-approximate $t'$-design (greater $t'$ indicates higher complexity), the projected ensemble $\mathcal{E}_{|\Psi\rangle}$ will form an $\varepsilon$-approximate $t$-design with probability $1-\delta$ for any positive $\varepsilon, \delta$, and integer $t
\geq 1$, given that $N_A$ and $N_B$ are sufficiently large. This provides an efficient scheme for generating random state ensembles from a single (complex) many-body state with projective measurements without having to sample from random quantum circuit  realizations, as in the conventional approach.

\vspace{8pt}
\emph{Magic states.}---The notion of magic states stems from the 
stabilizer formalism of fault-tolerant quantum computing. The latter was built upon a finite and non-universal set of unitary operations, the Clifford gates. Due to their non-universality, quantum circuits consisting of only Clifford operations can only produce a proper subset of all quantum states (from a fixed initial state). States that cannot be produced by Clifford circuits applied on initial computational basis states are called non-stabilizer states. Remarkably, it is known that by supplementing stabilizer operations with certain non-stabilizer states, one may achieve universal quantum computation.
The degree of non-stabilizerness of a state is typically referred to as its ``magic,'' and may be quantified by various measures~\cite{Veitch2014,Bravyi2016,Howard2017,Heinrich2019,Bravyi2019,Wang2020,Heimendahl2021, Haug2023scalablemeasures}. Several methods have been developed for numerically computing resource-theoretic measures of magic~\cite{Heinrich2019, Lami2023Non, Lami2024Unveiling, Tarabunga2024}, and the relationships between these measures and entanglement properties of quantum states have been explored~\cite{Tirrito2023, Turkeshi2023, Lami2024Quantum}. These concepts have found applications across a variety of areas, including classical simulation of quantum circuits with few non-stabilizer components~\cite{Bravyi2016, Bravyi2019, Heimendahl2021} and the study of measurement-induced phase transitions~\cite{Fux2023Entanglement, Bejan2023}. Of particular interest to our work is a recently developed family of measures of magic known as stabilizer entropies~\cite{Leone2022,Haug2023stabilizerentropies,Haug2023efficientSE,Haug2023MPS}, which have the advantage of being defined in an efficiently computable way rather than through a variational definition.
In particular, we will employ the stabilizer linear entropy from this family. 


More precisely, denote by $\mathcal{P}_n$ the set of all tensor products $P_1 \otimes \dots \otimes P_n$ where $P_i$ is a single qubit Pauli operator for $1\leq i \leq n$. Note that $\mathcal{P}_n$ is merely the set of all Pauli strings on $n$ systems and not the whole Pauli group, which includes global phase factors of $\pm 1, \pm i$. For any given state $|\psi\rangle$ over $n$ qubits and Pauli string $P\in \mathcal{P}_n$, define $\Xi_P(|\psi\rangle) \equiv \frac{1}{2^n} \langle \psi| P |\psi\rangle^2$, the (normalized) square of the expectation value of $P$ over state $|\psi\rangle$. From this, we may define the vector $\Xi(|\psi\rangle) \equiv (\Xi_P(|\psi\rangle))_{P\in \mathcal{P}_n}$, which has $4^n$ real components. The \textit{stabilizer linear entropy} is defined to be 
\begin{equation}\label{eq:magic}
    M_{\rm lin}(|\psi\rangle) = 1 - 2^n | \Xi(|\psi\rangle) |_2^2,
\end{equation}
where $| \cdot |_2$ is simply the Euclidean norm on $\mathbb{R}^{4^n}$. This measure satisfies the properties desired for good measures of nonstabilizerness from the point of view of resource theory~\cite{Leone2022} and offers the computational advantages for the purpose of the present work.

\vspace{8pt}
\emph{Connecting magic and randomness.}---We are now ready to establish the link between magic and randomness. Suppose that we generate an ensemble $\mathcal{E}_{|\Psi\rangle}$ from a state $|\Psi\rangle$ via the subsystem measurement introduced above. One might ask how the randomness of the resulting projected ensemble, as measured by distance $d^{(t)}(\mathcal{E}_{|\Psi\rangle})$ in~\eqref{eq:distance-def}, depends on the magic of state $|\Psi\rangle$, as quantified by the stabilizer linear entropy $M_{\rm lin}(|\Psi\rangle)$ in~\eqref{eq:magic}. It is worth stressing that ``magic'' is an operational notion defined with respect to the stabilizer formalism: states that are transformable by Clifford operations have the same magic value. The spirit lies behind other resource theories, such as entanglement: states transformable by local operation and classical communications are defined to have the same entanglement. Therefore, we treat magic as a property of a \textit{class} of states; the magic-randomness relationship is then established by averaging states with the same magic value, i.e., 
\begin{equation}\label{eq:Clifford-avg-def}
    \left\langle d^{(t)}(\mathcal{E}_{C|\Psi\rangle}) \right\rangle_{\mathcal{C}_N} \equiv \int_{C\in\mathcal{C}_N} dC~d^{(t)}(\mathcal{E}_{C|\Psi\rangle}).
\end{equation}
Note that in the above equation the average is equivalently performed over the Clifford group $\mathcal{C}_N$, rather than a general unitary group.
The task is then to determine the relationship between $\left\langle d^{(t)}(\mathcal{E}_{C|\Psi\rangle}) \right\rangle_{\mathcal{C}_N}$ and $M_{\rm lin}(|\Psi\rangle)$. This leads to the key result of this study.


\begin{theorem}\label{main-thm} 
Up to an error that decays with respect to $d_A$, the average randomness extracted from $|\Psi\rangle$ is described by
\begin{equation}\label{eq:main-thm-eq}
\begin{split}
    \left\langle [d_{\rm HS}^{(2)}(\mathcal{E}_{C|\Psi\rangle})]^2 \right\rangle_{\mathcal{C}_N} &= \alpha - \beta M_{\rm lin}(|\Psi\rangle) + \varepsilon,
\end{split}
\end{equation}
where $\alpha$ and $\beta$ are numerical factors that depend on both subsystem dimensions. 
\end{theorem}
Furthermore, we estimate the bound on the error $\varepsilon$ to be 
\begin{equation}\label{eq:error}
    \frac{|\varepsilon|}{\left\langle [d_{\rm HS}^{(2)}(\mathcal{E}_{C|\Psi\rangle})]^2 \right\rangle_{\mathcal{C}_N}} = \Theta\bigg(\frac{1}{d_A}\bigg),
\end{equation}
which implies that the error $\varepsilon$ decays exponentially faster in $N_A$ than the exact quantity $\langle [d_{\rm HS}^{(2)}(\mathcal{E}_{C|\Psi\rangle})]^2 \rangle_{\mathcal{C}_N}$. This error term comes from estimating the average of a quotient by separately averaging the numerator and denominator. Proof of the theorem and derivation of the error estimate are delegated to the Supplementary Material (SM). We add that we consider the average of the square of $d_{\rm HS}^{(2)}(\mathcal{E}_{C|\Psi\rangle})$ rather than the Hilbert-Schmidt distance itself because doing so allows us to avoid the difficulty of averaging the square root of the operator trace, and hence deduce a rigorous result. The average-of-squares $\langle [d_{\rm HS}^{(2)}(\mathcal{E}_{C|\Psi\rangle})]^2\rangle_{\mathcal{C}_N}$ is expected to approximate the square-of-average $\langle d_{\rm HS}^{(2)}(\mathcal{E}_{C|\Psi\rangle})\rangle^2_{\mathcal{C}_N}$ for large systems due to concentration of measure, with an error suppressed exponentially fast with the system size. Rigorously proving this conjecture is challenging due to the complexity of evaluating higher moments of the Clifford group. Instead, we provide supporting numerical evidence in the SM.

While previous work has analyzed the magic content of states in the context of quantum typicality~\cite{Gu2023,Turkeshi2023Pauli}, this theorem implies a direct connection between the randomness one may extract from quantum states and their magic. It gives a new meaning to the stabilizer linear entropy of an arbitrary state in terms of the extractable randomness from that state. We note that the ``stabilizer purification'' framework of~\cite{Bejan2023} may provide a useful approach for analyzing the mechanism by which this relationship between randomness and magic arises. Finally, we add that since the coefficients tend to zero as $d_A, d_B\to\infty$, Theorem~\ref{main-thm} confirms that as one increases the size of the subsystem measurement and of the surviving systems, the values of $d_{\rm HS}^{(2)}(\mathcal{E}_{C|\Psi\rangle})$ tend to zero, i.e., the randomness becomes near perfect. 

\begin{figure}[t!]
    \centering
    \includegraphics[width=\columnwidth]{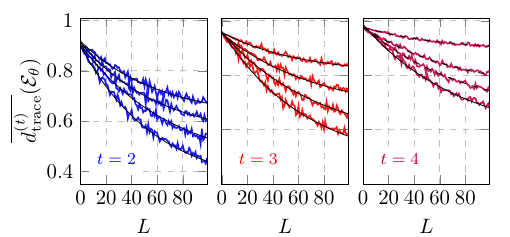}
    \caption{We numerically estimate $\overline{d^{(t)}_{\mathrm{trace}}}(\mathcal{E}_\theta)$ (colored solid lines) with $N_A = 2$ and $N_B = 4$ and provide an exponential fit (black solid line). For each circuit depth, the numerical average is taken over 100 realizations of a random Clifford circuit of the given depth initialized to $|\Psi(\theta)\rangle$. We plot $\overline{d^{(t)}_{\mathrm{trace}}}(\mathcal{E}_\theta)$ as a function of circuit depth assuming initial states $|\Psi(\theta)\rangle$ of magic value $M_{\mathrm{lin}}(|\Psi(\theta)\rangle) = 0, 0.25, 0.5, 0.75$, with lower curves corresponding to higher magic values. }    
    \label{fig:decay}
\end{figure}

\vspace{8pt}
\emph{Simulations.}--- Due to the rapid growth in the size of the Clifford group $\mathcal{C}_N$ with respect to $N$, it is difficult to numerically compute our quantifier of randomness $\left\langle d^{(t)}(\mathcal{E}_{C|\Psi\rangle}) \right\rangle_{\mathcal{C}_N}$.
To circumvent this issue, we estimate $\left\langle d^{(t)}(\mathcal{E}_{|C\Psi\rangle}) \right\rangle_{\mathcal{C}_N}$ by averaging across random Clifford circuits of restricted length and extrapolating the results. Furthermore, another advantage of considering simulations of random Clifford circuits together with subsystem measurements is that they represent a physical quantum information processing task in which nonstabilizerness can affect the formation of approximate state designs. 

We consider an $N$-qubit product state $|\Psi(\theta)\rangle = 2^{-N/2} (|0\rangle + e^{i\theta} |1\rangle)^{\otimes N}$, which has a magic value $M_{\text{lin}}(|\Psi(\theta)\rangle)=1-(1+\cos^4{\theta}+\sin^4{\theta})^N/d$~\cite{Tirrito2023}. The state $|\Psi(\theta)\rangle$ is evolved according to a random Clifford circuit 
$C_L = \prod_{i = 1}^{L} U_i$ of depth $L$. Each layer $U_i$ of this circuit (Fig.~\ref{fig:setup}, right) acts on two registers randomly sampled from the $N$ total registers and is itself randomly selected from the two-qubit gate set $\{\text{CNOT}, H\otimes I, S\otimes I$, $I\otimes H$, $I \otimes S$\}, where $H$ is the Hadamard gate and $S = \text{diag}(1, \ e^{i\pi/4})$ is the phase shift gate. The projected ensemble $\mathcal{E}_{C_L|\Psi(\theta)\rangle}$ over the $A$ subsystem is then obtained by simulating measurements on the state $C_L|\Psi(\theta)\rangle$ over the $B$ subsystem. 
Define $\overline{d_{\rm trace}^{(t)}}(\mathcal{E}_\theta)$ to be the average of $d_{\rm trace}^{(t)}(\mathcal{E}_{C_L|\Psi(\theta)\rangle})/2$ over all possible Clifford circuits $C_L$ of depth $L$ composed of two qubit unitaries in the previously described manner. 
Here, we insert a normalization factor of $1/2$ to ensure that $\overline{d_{\rm trace}^{(t)}}(\mathcal{E}_\theta)$ lies in $[0,1]$ for illustrative purposes. We emphasize that the bar is used to indicate an average over all Clifford circuits of a specified length, rather than the entire Clifford group as in~\eqref{eq:Clifford-avg-def}. A numerical estimate for the trace distance $\overline{d_{\text{trace}}^{(t)}}(\mathcal{E}_\theta)$ is obtained by averaging over many realizations of a random Clifford circuit. 
By varying the depth of the simulated circuit, we obtain data for $\overline{d_{\text{trace}}^{(t)}}(\mathcal{E}_\theta)$ as a function of circuit depth $L$. 

\begin{figure}[t!]
    \centering
    \includegraphics[width=\columnwidth]{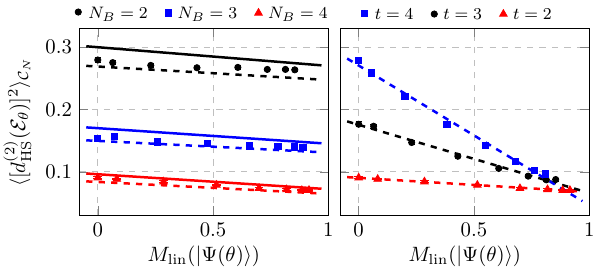}
    \caption{Numerical estimates for the average extracted second degree randomness $\langle [d_{\mathrm{HS}}^{(2)} (\mathcal{E}_\theta)]^2\rangle_{C_N}$ of the initial state $|\Psi(\theta)\rangle$ as a function of its magic value $M_{\mathrm{lin}}(|\Psi(\theta)\rangle)$ (data points) for various system sizes and moments $t$. The numerical estimates were obtained by averaging over $100$ realizations per circuit depth of random Clifford circuits with depth ranging from 1 to 200, and extrapolating to larger depth via an exponential fit. Left: Solid lines are the analytic approximation~\eqref{eq:main-thm-eq}. The deviation from the analytic approximation (see the SM) by an amount given by our error estimate \eqref{eq:error} is shown in dashed line. Error bars indicating the standard error of the asymptotic fitting parameter were plotted and are obscured by the data points. All results were obtained for $N_A = 5$. Right: Data were obtained for $N_{A(B)} = 2(4)$ at $t = 4$, $N_{A(B)} = 4(3)$ at $t = 3$, and $N_{A(B)} = 5(4)$ at $t = 2$. Dashed lines represent the best linear fits.}
    \label{fig:linear}
\end{figure}

Numerically, we see that the average distance $\overline{d_{\text{trace}}^{(2)}}(\mathcal{E}_\theta)$ of the post-measurement ensemble decays exponentially with respect to the circuit depth $L$, with values of $\theta$ corresponding to a higher magic value producing a faster decay (Fig.~\ref{fig:decay}). The variation in the decay curves shows that the magic of many-body quantum states influences the degree of randomness that may be extracted from them via local measurements. Furthermore, it was observed that the asymptotic value of the decay in average trace distance falls exponentially with respect to $N_B$ for fixed $N_A$ and magic, while it decreases linearly with respect to $M_{\text{lin}}(|\Psi(\theta)\rangle)$ for fixed $N_A$ and $N_B$. 
The latter is consistent with Theorem~\ref{main-thm} (see~\footnote{At great depth, the random circuit average $\protect \overline {d_{\protect \rm trace}^{(2)}}(\protect \mathcal {E}_{\theta })$ approximates the Clifford group average $\protect \frac {1}{2}\langle d_{\protect \text {trace}}^{(2)}(\protect \mathcal {E}_{C|\Psi (\theta )\rangle })\rangle _{C\in \protect \mathcal {C}_N}$, whose value we can expect to approximate $\protect \sqrt {\langle [\protect \frac {1}{2}d_{\protect \text {trace}}^{(2)}(\protect \mathcal {E}_{C|\Psi (\theta )\rangle })]^2\rangle _{C\in \protect \mathcal {C}_N}}$ due to measure concentration. Furthermore, we may then use the inequality $d_{\protect \rm trace}(\cdot ) \leq \protect \sqrt {d} d_{\protect \rm HS}(\cdot )$ to estimate $\langle d_{\protect \text {trace}}^{(2)}(\protect \mathcal {E}_{C|\Psi (\theta )\rangle })\rangle _{C\in \protect \mathcal {C}_N}$ by a constant multiple of the square root of the RHS of~\protect \eqref {eq:main-thm-eq}. Due to the small slope in the linear function~\protect \eqref {eq:main-thm-eq}, we can then
deduce that the asymptotic values of $\protect \overline {d_{\protect \text {trace}}^{(2)}}(\protect \mathcal {E}_\theta )$ should decrease approximately linearly with respect to $M_{\protect \text {lin}}(|\Psi (\theta )\rangle )$ (See Fig.~\ref {fig:linear}). Finally, we note the scaling factor of the exponential decay appears to be roughly constant with respect to $M_{\protect \text {lin}}(|\Psi (\theta )\rangle )$ when holding $N_A$ and $N_B$ fixed.}).

The exponential decay of the $\overline{d_{\rm trace}^{(t)}}(\mathcal{E}_\theta)$ with respect to circuit depth may be seen as a general property of random circuits: For many random circuit configurations, the behavior of $d^{(t)}(\mathcal{E})$ has been studied~\cite{BHH2016, Harrow2009}, where in this case $\mathcal{E}$ is the ensemble of all possible states that may be evolved from a fixed initial state of the random circuit. Generally, any positive value $\varepsilon$ of $d^{(t)}(\mathcal{E})$ may be achieved provided the circuit depth $L$ scales at least linearly with respect to $t$, number of registers $N$, and $\log(1/\varepsilon)$. The linear growth of this threshold with respect to $\log(1/\varepsilon)$ suggests that $d^{(t)}(\mathcal{E})$ decays exponentially with respect to circuit depth in these random circuit paradigms. Since the measurements in our scheme introduce an extra degree of randomness, our value of $\overline{d^{(t)}}(\mathcal{E}_\theta)$ should decay at least exponentially fast, as was observed numerically. Furthermore, the exponential decay of the asymptotic values with respect to $N_B$ follow from Theorem 2 of  Ref.~\cite{Cotler2023}, where the bounds on the value of $N_B$ necessary to produce a post-measurement ensemble $\mathcal{E}$ with a value of $d_{\rm trace}^{(t)}(\mathcal{E}) \leq \varepsilon$ require that $N_B$ scale exponentially faster than $\log (1/ \varepsilon)$. 

We conclude our numerical study with a demonstration of Theorem~\ref{main-thm}. As before, we define $\overline{(d_{\rm HS}^{(t)})^2}(\mathcal{E}_\theta)$ to be the average of $[d_{\rm HS}^{(t)}(\mathcal{E}_{C_L|\Psi(\theta)\rangle})]^2$ over all possible Clifford circuits $C_L$ of depth $L$ composed according to the aformentioned two-qubit gate prescription. By numerically estimating $\overline{(d_{\rm HS}^{(t)})^2}(\mathcal{E}_\theta)$ with many iterations of a finite-depth Clifford circuit, we obtain data for the exponential decay of $\overline{(d_{\rm HS}^{(t)})^2}(\mathcal{E}_\theta)$ as a function of circuit depth. Fitting this data to an exponential decay function, we are able to extrapolate to infinite depth so that we obtain an approximation for $\langle [d_{\rm HS}^{(t)}(\mathcal{E}_{C|\Psi(\theta)\rangle})]^2\rangle_{\mathcal{C}_N}$, which we abbreviate to $\langle [d_{\rm HS}^{(t)}(\mathcal{E}_{\theta})]^2\rangle_{\mathcal{C}_N}$. By repeating this process for various angles $\theta$, we obtain $\langle [d_{\rm HS}^{(t)}(\mathcal{E}_\theta)]^2\rangle_{\mathcal{C}_N}$ as a function of $M_{\rm lin}(|\Psi(\theta)\rangle)$, as seen in Fig.~\ref{fig:linear} for $t = 2$. Note the linear nature of the relationship between $\langle [d_{\rm HS}^{(2)}(\mathcal{E}_\theta )]^2\rangle_{\mathcal{C}_N}$ and magic. Furthermore, the numerically simulated values for $\langle [d_{\rm HS}^{(2)}(\mathcal{E}_\theta)]^2\rangle_{\mathcal{C}_N}$ fall below those of the plotted analytical approximation~\eqref{eq:main-thm-eq}. We have numerically verified that these data fall within the error threshold given by~\eqref{eq:error}. Further analysis shows that we can expect the deviation between the values of $\langle [d_{\rm HS}^{(2)}(\mathcal{E}_\theta)]^2\rangle_{\mathcal{C}_N}$ and its analytical approximation to be uniformly negative (see the SM). Lastly, our numerical simulations (Fig.~\ref{fig:linear}, right) suggest that the linear relationship between $\langle [d_{\rm HS}^{(t)}(\mathcal{E}_\theta )]^2\rangle_{\mathcal{C}_N}$ and $M_{\text{lin}}(|\Psi(\theta)\rangle)$ holds for higher values of $t$. Moreover, the enhancement of randomness by magic is more prominent for larger values of $t$, further justifying the efficiency of the protocol for random state generation.  

\vspace{8pt}
\emph{Discussion.}---
One of the most notable aspects of this study is the transformative influence of the magic inherent in the initial state on the resulting randomness in the projected ensemble. It is crucial to stress that the Clifford group constitutes a quantum unitary $3$-design~\cite{Zhu2017}. This implies that the produced ensemble (of the total system before projective measurement) by random Clifford circuits forms a quantum state $3$-design as well, irrespective with the initial input state. Consequently, one might naturally anticipate that the projected ensemble's randomness level should be independent of the initial state, contrary to our observation. 

The observed augmentation of the randomness level of the projected ensemble due to the amount of magic reflects internal correlations among subsystems within each \emph{single} quantum state of the ensemble. These internal correlations are not manifest in the quantum state without subsystem projective measurements; they also go \emph{beyond} what conventional metrics can capture. For example, entanglement of bipartite pure quantum states are completely determined by the reduced density matrix of the subsystems~\cite{Horodecki2009}, representing a first-moment average of the projected ensemble [$t=1$ in~\eqref{eq:moment}]. The connection between magic and randomness thus suggests a novel approach for characterizing and harnessing correlations in magic states. 

Simultaneously, leveraging the benefits of high-order quantum state designs necessitates multiple copies of each states within the ensemble. This poses a significant constraint on the precision of random state generation protocols, challenging both the conventional method such as sampling random unitary circuits and the recently proposed projected ensemble approach. The strategy proposed in this work, involving the use of Clifford circuits, marks a crucial first step toward developing a fault tolerant protocol for random state generation. 

\vspace{8pt}
Acknowledgement.---The authors thank Lorenzo Leone for helpful discussions. This work was supported in part by the U.S. Department of Energy, Office of Science, Office of Advanced Scientific Computing Research, through the Quantum Internet to Accelerate Scientific Discovery Program, and in part by the LDRD program at Los Alamos. C.V. also acknowledges support from the Quantum Computing Summer School at Los Alamos National Laboratory.

\bibliography{references}

\clearpage
\appendix

\setcounter{page}{1}
\renewcommand\thefigure{\thesection\arabic{figure}}
\setcounter{figure}{0} 

\onecolumngrid

\begin{center}
\large{ Supplemental Material for \\ Extracting randomness from quantum `magic'
}
\end{center}

Here we present the proof of Theorem 1 in the main text.

\section{Overall picture}

Fix $|\psi\rangle \in \mathcal{H}_A \otimes \mathcal{H}_B$ and define $P_i \equiv I_A \otimes |i\rangle \langle i|_B$ for each bit string $i\in \{0,1\}^{N_B}$. For any Clifford unitary $C\in \mathcal{C}_N$, we define $\rho_C^{(2)}$ to be the second moment of the post-measurement ensemble obtained from the generator state $C|\psi\rangle$:
\begin{equation}
        \rho_C^{(2)} = \sum_{i\in\{0,1\}^{N_B}} \frac{\text{Tr}_B[P_iC\psi C^{\dagger}P_i]^{\otimes 2}}{\text{Tr}[P_i C \psi C^\dagger]}.
\end{equation}
We wish to calculate 
\begin{equation}\label{eq:goal}
\begin{split}
    \langle \|\rho_C^{(2)} - \rho_H^{(2)}\|_2^2 \rangle_{\mathcal{C}_N} &=
    \langle \text{Tr}[(\rho_C^{(2)} - \rho_H^{(2)})^2] \rangle_{\mathcal{C}_N} \\ &=
    \underbrace{\langle \text{Tr}[(\rho_C^{(2)})^2]\rangle_{\mathcal{C}_N}}_{\text{(I)}}  - 2\underbrace{\langle\text{Tr}[\rho_C^{(2)}\rho_H^{(2)}]\rangle_{\mathcal{C}_N}}_{\text{(II)}} + \underbrace{\text{Tr}[(\rho_H^{(2)})^2]}_{\text{(III)}} 
\end{split}
\end{equation} Define $d = 2^N$. 
Let $\mathcal{P}_N$ be the set of all tensor products of Pauli operators over $N$ qubits, and for all $P\in \mathcal{P}_N$ define $\Xi_P(|\psi\rangle) = d^{-1}\langle\psi|P|\psi\rangle^2$. Then construct the $4^N$-component vector $\Xi(|\psi\rangle) = (\Xi_P(|\psi\rangle))_{P\in \mathcal{P}_N}$. The \textit{stabilizer linear entropy} is defined as
\begin{equation}
    M_{\rm lin}(|\psi\rangle) = 1- d\|\Xi(|\psi\rangle)\|_2^2.
\end{equation}
Our ultimate goal is to express~\eqref{eq:goal} as a function of the nonstabilizerness of $|\psi\rangle$. The metric of nonstabilizerness that we will use is the stabilizer linear entropy. In what follows, we assume that $d_A \equiv 2^{N_A}, d_B \equiv 2^{N_B} \gg 1$. 

\subsection{Evaluation of Term (I)}

We will evaluate each term in~\eqref{eq:goal} individually, starting with term I. We have
\begin{equation}\label{eq:Tr-rho2-calc}
\begin{split}
    \langle \text{Tr}[(\rho_C^{(2)})^2] \rangle_{\mathcal{C}_N} &= 
    \bigg\langle
    \text{Tr}\bigg\{\sum_{i,j \in \{0, 1\}^{N_B}} \frac{\text{Tr}_B[P_i C \psi C^\dagger P_i]^{\otimes 2}}{\text{Tr}[P_i C\psi C^\dagger]} \frac{\text{Tr}_B[P_jC\psi C^\dagger P_j]^{\otimes 2}}{\text{Tr}[P_j  C\psi C^\dagger]}\bigg\}\bigg\rangle_{\mathcal{C}_N} \\
    &=
    \sum_{i,j \in \{0,1\}^{N_B}} \bigg\langle \frac{\text{Tr}[S_A([P_iC\psi C^\dagger P_i]^{\otimes 2}\otimes [P_j C\psi C^\dagger P_j]^{\otimes 2})]}{\text{Tr}[(P_iC\psi C^\dagger) \otimes (P_j C\psi C^\dagger)]} \bigg\rangle_{\mathcal{C}_N} \\
    &= \sum_{i,j \in \{0,1\}^{N_B}} \bigg\langle \frac{\text{Tr}[S_A([P_i^{\otimes 2}\otimes P_j^{\otimes 2}][C^{\otimes 4}\psi^{\otimes 4} C^{\dagger \otimes 4}][P_i^{\otimes 2}\otimes P_j^{\otimes 2}])]}
    {\text{Tr}[(P_i\otimes P_j)(C^{\otimes 2}\psi^{\otimes 2} C^{\dagger \otimes 2})]} \bigg\rangle_{\mathcal{C}_N} \\
    &\approx \sum_{i,j \in \{0,1\}^{N_B}} \frac{\langle \text{Tr}[S_A([P_i^{\otimes 2}\otimes P_j^{\otimes 2}][C^{\otimes 4}\psi^{\otimes 4} C^{\dagger \otimes 4}][P_i^{\otimes 2}\otimes P_j^{\otimes 2}])]\rangle_{\mathcal{C}_N}}
    {\langle \text{Tr}[(P_i\otimes P_j)(C^{\otimes 2}\psi^{\otimes 2} C^{\dagger \otimes 2})]\rangle_{\mathcal{C}_N}} \\
    &= \sum_{i,j \in \{0,1\}^{N_B}} \frac{ \text{Tr}[S_A([P_i^{\otimes 2}\otimes P_j^{\otimes 2}]\langle C^{\otimes 4}\psi^{\otimes 4} C^{\dagger \otimes 4}\rangle_{\mathcal{C}_N}[P_i^{\otimes 2}\otimes P_j^{\otimes 2}])]}
    {\text{Tr}[(P_i\otimes P_j)\langle C^{\otimes 2} \psi^{\otimes 2} C^{\dagger \otimes 2}\rangle_{\mathcal{C}_N}]},
\end{split}
\end{equation}
where the numerators in the above expressions are a function of operators over a fourfold copy of the Hilbert space $\mathcal{H}_A \otimes \mathcal{H}_B$, and $S_A$ is the swap operator acting on the $A$ parts of these Hilbert spaces across the partition $12|34$. The justification for the approximation on the fourth line is given below in the subsection titled ``Mean Quotient Approximation for Term I.''

\subsubsection{Evaluation of Denominator}

Let us calculate the denominator of the expression in~\eqref{eq:Tr-rho2-calc} first. Since the uniformly-weighted Clifford group forms a 2-design, we automatically have
\begin{equation}
    \langle C^{\otimes 2}\psi^{\otimes 2}C^{\dagger \otimes 2} \rangle_{\mathcal{C}_N} = \frac{\Pi_2^{sym}}{\text{Tr}[\Pi_2^{sym}]},
\end{equation}
where for all $k>1$, we define $\Pi_k^{sym} = \sum_{\rho \in S_k} T_\rho$, with $T_\rho$ being the operator over $(\mathcal{H}_A \otimes \mathcal{H}_B)^{\otimes k}$ that permutes the $k$ tensor factors in this space according to the permutation $\rho$. The operator $\Pi_k^{sym}$ is the unnormalized projector onto the symmetric subspace of $(\mathcal{H}_A \otimes \mathcal{H}_B)^{\otimes k}$, and has trace given by $\text{Tr} [\Pi_k^{sym}] = d(d+1) \dots (d+k - 1)$. Note that permutations over $(\mathcal{H}_A\otimes\mathcal{H}_B)^{\otimes k}$ factor as $T_\rho = T_\rho^A \otimes T_\rho^B$, where $T_\rho^A$ and $T_\rho^B$ are the operators corresponding to the action of the permutation $\rho\in S_k$ over $\mathcal{H}_A^{\otimes k}$ and $\mathcal{H}_B^{\otimes k}$, respectively. We then have
\begin{equation}\label{eq:term1-denom}
    \begin{split}
        \text{Tr}[(P_i\otimes P_j)\langle C^{\otimes 2} \psi^{\otimes 2} C^{\dagger \otimes 2}\rangle_{\mathcal{C}_N}] &= \sum_{\rho \in S_2}
        \frac{\text{Tr}[([I_A \otimes |i\rangle\langle i|_B]\otimes[I_A \otimes |j\rangle\langle j|_B]) (T_\rho^A \otimes T_\rho^B)]}{d_Ad_B(d_Ad_B + 1)}\\
        &= \sum_{\rho \in S_2} \frac{\text{Tr}[T_\rho^A\otimes ([|i\rangle\langle i|_B \otimes |j\rangle\langle j|_B ]T_\rho^B)]}{d_Ad_B(d_Ad_B + 1)} \\
        &= \sum_{\rho \in S_2} \frac{\text{Tr}[T_\rho^A]\langle i j|_B T_\rho^B |ij\rangle_B}{d_Ad_B(d_Ad_B + 1)} \\
        &=\begin{cases} 
              \frac{\text{Tr}[T_\text{Id}^A]\langle i j|_B T_\text{Id}^B |i j\rangle_B}{d_Ad_B(d_Ad_B + 1)}, & i\neq j \\
              \sum_{\rho\in S_2} \frac{\text{Tr}[T_\rho^A]}{d_Ad_B(d_Ad_B + 1)}, & i = j 
        \end{cases} \\
        &= \begin{cases} 
              \frac{d_A}{d_B(d_Ad_B + 1)} & i\neq j \\
              \frac{d_A + 1}{d_B(d_Ad_B + 1)} & i = j 
        \end{cases} 
    \end{split}
\end{equation}

\subsubsection{Rewriting the Numerator}

We proceed to calculate the numerator of~\eqref{eq:Tr-rho2-calc}. By a result of~\cite{Leone2021, Zhu2016}, we have that
\begin{equation}\label{eq:clifford-4th-average}
\begin{split}
    \langle C^{\otimes 4} \psi^{\otimes 4} C^{\dagger \otimes 4} \rangle_{\mathcal{C}_N} &= aQ\Pi_4^{\rm sym} + b\Pi_4^{\rm sym}, 
\end{split}
\end{equation}
where 
\begin{equation}\label{eq:alpha-beta-defs}
\begin{split}
    &a = \frac{\| \Xi(|\psi\rangle) \|_2^2}{4(d_Ad_B + 1)(d_Ad_B + 2)} - b\\
    &b = \frac{1 - \| \Xi(|\psi\rangle) \|_2^2}{(d_A^2 d_B^2 - 1)(d_Ad_B + 2)(d_Ad_B + 4)}\\
    &Q = \frac{1}{d_A^2 d_B^2} \sum_{P\in \mathcal{P}_N} P^{\otimes 4}.
\end{split}
\end{equation}
In this instance, $\mathcal{P}_N$ simply denotes the set of tensor products of Pauli operators of length $N$, \textit{not} the entire Pauli group on $N$ qubits (i.e., it does not include phase prefactors). We briefly mention that our approach of averaging over the Clifford group by using \eqref{eq:clifford-4th-average} to relate magic to another quantity of interest resembles the techniques used in~\cite{Tirrito2023, Turkeshi2023}. 
The numerator of the expression in~\eqref{eq:Tr-rho2-calc} now becomes
\begin{equation}\label{eq:num-expr}
\begin{split}
    \text{Tr}[S_A([P_i^{\otimes 2}\otimes P_j^{\otimes 2}]\langle C^{\otimes 4}\psi^{\otimes 4} C^{\dagger \otimes 4}\rangle_{\mathcal{C}_N}[P_i^{\otimes 2}\otimes P_j^{\otimes 2}])] &=
    a\text{Tr}[S_A([P_i^{\otimes 2}\otimes P_j^{\otimes 2}]Q\Pi_4^{\rm sym}[P_i^{\otimes 2}\otimes P_j^{\otimes 2}])] \\ &+
    b\text{Tr}[S_A([P_i^{\otimes 2}\otimes P_j^{\otimes 2}]\Pi_4^{\rm sym}[P_i^{\otimes 2}\otimes P_j^{\otimes 2}])]. 
\end{split}
\end{equation}

\subsubsection{Evaluation of Second Numerator Term}
We evaluate the latter term of the numerator~\eqref{eq:num-expr} first. If $i\neq j$, one has
\begin{equation}
    \begin{split}
        \text{Tr}[S_A([P_i^{\otimes 2}\otimes P_j^{\otimes 2}]\Pi_4^{\rm sym}[P_i^{\otimes 2}\otimes P_j^{\otimes 2}])] &= 
        \sum_{\rho\in S_4} \text{Tr}[S_A([P_i^{\otimes 2}\otimes P_j^{\otimes 2}]T_\rho [P_i^{\otimes 2}\otimes P_j^{\otimes 2}])] \\
        &=
        \sum_{\rho\in S_4} \text{Tr}[S_A([(I_A\otimes |i\rangle\langle i|_B)^{\otimes 2}\otimes (I_A\otimes |j\rangle\langle j|_B)^{\otimes 2}]
        \\&\ \ \ \ \ \ \ \ \ \ \ \ \times
        [T_\rho^A \otimes T_\rho^B][(I_A\otimes |i\rangle\langle i|_B)^{\otimes 2}\otimes (I_A\otimes |j\rangle\langle j|_B)^{\otimes 2}])] \\
        &= 
        \sum_{\rho\in S_4} \text{Tr}[S_A(T_\rho^A \otimes |iijj\rangle\langle iijj|_BT_\rho^B|iijj\rangle\langle iijj|_B
        )] \\
        &=
        \sum_{\rho\in \{\text{Id}, (12), (34), (12)(34)\}} \text{Tr}[S_A(T_\rho^A \otimes |iijj\rangle\langle iijj|_B)] \\
        &=
        \sum_{\rho, \sigma\in S_2} \text{Tr}[(S_A [T_\rho^{A, 12} \otimes T_\sigma^{A, 34}]) \otimes |iijj\rangle\langle iijj|_B] \\
        &=
        \sum_{\rho, \sigma\in S_2} \text{Tr}[S_A(T_\rho^{A,12}\otimes T_\sigma^{A,34})] \\
        &=
        \text{Tr}[S_A(\Pi_{2, A}^{sym} \otimes \Pi_{2, A}^{sym})] \\
        &= 
        \text{Tr}[(\Pi_{2,A}^{sym})^2] \\
        &=
        \text{Tr}[2(\Pi_{2,A}^{sym})] \\
        &=2d_A(d_A + 1).
    \end{split}
\end{equation}
The fourth equality follows from the fact that the only permutations that permute the tensor factors in the state $|iijj\rangle$ such that it remains unchanged are ${\rm Id}, (12), (34)$, and $(12)(34)$. In the fifth equality, $T_\rho^{A,12}$ indicates the operator corresponding to the action of $\rho\in S_2$ on the first and second copies of the $A$ system, and $T_\sigma^{A,34}$ indicates the operator corresponding to the action of $\sigma\in S_2$ on the third and fourth copies of the $A$ system. This equality is easily seen by writing $T_{(12)\in S_4} = T_{(12)\in S_2}^{A, 12}\otimes T_{\text{Id}\in S_2}^{A, 34}$, $T_{(34)\in S_4} = T_{\text{Id}\in S_2}^{A, 12}\otimes T_{(12)\in S_2}^{A, 34}$, $T_{(12)(34)\in S_4}^A = T_{(12)\in S_2}^{A,12} \otimes T_{(12)\in S_2}^{A,34}$. The second-to-last equality follows simply from the fact that $(\text{Id} + (12))^2 = 2(\text{Id} + (12))$. 

On the other hand, if $i  =  j$, we can reuse some of the preceding arguments to get
\begin{equation}
    \begin{split}
        \text{Tr}[S_A([P_i^{\otimes 2}\otimes P_j^{\otimes 2}]\Pi_4^{\rm sym}[P_i^{\otimes 2}\otimes P_j^{\otimes 2}])] &= 
        \sum_{\rho\in S_4} \text{Tr}[S_A(T_\rho^A \otimes |iiii\rangle\langle iiii|_BT_\rho^B|iiii\rangle\langle iiii|_B
        )] \\ &=
        \sum_{\rho\in S_4} \text{Tr}[S_A T_\rho^A] \\
        &=
        \sum_{\rho\in S_4} \text{Tr}[T_{(13)(24)}^A T_\rho^A] \\
        &=
        \sum_{\rho\in S_4} \text{Tr}[T_\rho^A] \\
        &=
        \text{Tr}[\Pi_{4,A}^{sym}] \\
        &= d_A(d_A+1)(d_A + 2)(d_A + 3).
    \end{split}
\end{equation}
We can summarize the preceding information by writing
\begin{equation}\label{eq:num-second-term-final-expression}
    \text{Tr}[S_A([P_i^{\otimes 2}\otimes P_j^{\otimes 2}]\Pi_4^{\rm sym}[P_i^{\otimes 2}\otimes P_j^{\otimes 2}])] =
    \begin{cases}
        2d_A(d_A + 1) & i \neq j \\
        d_A(d_A+1)(d_A + 2)(d_A + 3) & i = j
    \end{cases}
\end{equation}

\subsubsection{Evaluation of First Numerator Term}
We proceed to evaluate the first term of the numerator~\eqref{eq:num-expr}. We first assume $i\neq j$. We have
\begin{equation}\label{eq:num-former-term}
    \begin{split}
        \text{Tr}[S_A([P_i^{\otimes 2}\otimes P_j^{\otimes 2}]Q\Pi_4^{\rm sym}[P_i^{\otimes 2}\otimes P_j^{\otimes 2}])] &=
        \sum_{\rho\in S_4} \text{Tr}[S_A([(I_A\otimes |i\rangle\langle i|_B)^{\otimes 2}\otimes (I_A\otimes |j\rangle\langle j|_B)^{\otimes 2}](Q_A\otimes Q_B)\\
        &\ \ \ \ \ \ \ \ \ \ \ \ \times (T_\rho^A\otimes T_\rho^B)[(I_A\otimes |i\rangle\langle i|_B)^{\otimes 2}\otimes (I_A\otimes |j\rangle\langle j|_B)^{\otimes 2}])] \\
        &= \sum_{\rho \in S_4} \text{Tr}[(S_AQ_AT_\rho^A)\otimes(|iijj\rangle\langle iijj|_B Q_B T_\rho^B |iijj\rangle\langle iijj|_B)] \\ 
        &= \sum_{\rho \in S_4} \text{Tr}[S_AQ_AT_\rho^A]\langle iijj|Q_B T_\rho^B|iijj\rangle
    \end{split}
\end{equation} 

The calculation of the matrix element $\langle iijj|Q_B T_\rho^B|iijj\rangle$ requires some special care. By writing $|Pk\rangle \equiv P|k\rangle$ for $k\in \{0,1\}^{N_B}$, we have
\begin{equation}\label{eq:mat-el-calc}
\begin{split}
    \langle iijj|Q_B T_\rho^B|iijj\rangle &=
    \frac{1}{d_B^2} \sum_{P\in \mathcal{P}_{N_B}} \langle iijj|P^{\otimes 4} T_\rho^B |iijj\rangle \\
    &= 
    \frac{1}{d_B^2} \sum_{P\in \mathcal{P}_{N_B}} \langle Pi, Pi, Pj, Pj| T_\rho^B |iijj\rangle.
\end{split}
\end{equation}
Note that if $\rho\notin \{\text{Id}, (12), (34), (12)(34), (13)(24), (14)(23), (1324), (1423)\}$, then the above matrix element is automatically zero. The reasoning for this is as follows. All other permutations transform the ket $|iijj\rangle$ into either 
$|jiji\rangle$, $|ijij\rangle$, $|jiij\rangle$, or $|ijji\rangle$. Furthermore, since $P$ is simply a Pauli string on $N_B$ qubits, it transforms $|i\rangle$ into another bit string (up to a phase) on $N_B$ qubits. Thus,
the overlap of $| Pi, Pi, Pj, Pj\rangle$ with  $|ijij\rangle$, $|jiji\rangle$, $|jiij\rangle$, and $|ijji\rangle$ is zero, for if it were not, our assumption that $i\neq j$ would be contradicted. 
Now suppose that $\rho \in \{\text{Id}, (12), (34), (12)(34)\}$. Then~\eqref{eq:mat-el-calc} becomes
\begin{equation}\label{eq:invariant-perms-mat-el}
\begin{split}
    \langle iijj|Q_B T_\rho^B|iijj\rangle &= 
    \frac{1}{d_B^2} \sum_{P\in \mathcal{P}_{N_B}} \langle Pi, Pi, Pj, Pj|iijj\rangle.
\end{split}
\end{equation}
Let us identify the Pauli strings $P$ for which the overlap $\langle Pi, Pi, Pj, Pj|iijj\rangle$ is nonzero. If this overlap is nonzero, then $P$ must be a Pauli string consisting of solely $I$ and $Z$ operators. Otherwise, it would flip one of the bits when it acts on the vector $|i\rangle$ or $|j\rangle$, which would render the overlap equal to zero. Furthermore, \textit{all} such $I, Z$-strings yield $\langle Pi, Pi, Pj, Pj|iijj\rangle = 1$ because such strings leave the bit strings unchanged except for a possible phase factor of $-1$, which would cancel out due to the double pairing of like vectors in the transformed state $|Pi, Pi, Pj, Pj\rangle$. Since there are $2^{N_B}$ Pauli strings consisting solely of $I$ and $Z$ operators, it follows that there are $2^{N_B}$ Pauli strings $P$ for which $\langle Pi, Pi, Pj, Pj|iijj\rangle = 1$, and for all others this overlap is zero. Hence,~\eqref{eq:invariant-perms-mat-el} becomes
\begin{equation}
    \langle iijj|Q_BT_\rho^B|iijj\rangle = \frac{1}{d_B^2} \cdot 2^{N_B}  = \frac{1}{d_B}
\end{equation}
if $\rho \in \{\text{Id}, (12), (34), (12)(34)\}$. Now assume $\rho \in \{(13)(24), (14)(23), (1324), (1423)\}$. In this case,~\eqref{eq:mat-el-calc} becomes
\begin{equation}\label{eq:flipping-perms-mat-el}
    \langle iijj|Q_BT_\rho^B|iijj\rangle = \frac{1}{d_B^2} \sum_{P\in \mathcal{P}_{N_B}} \langle Pi, Pi, Pj, Pj|jjii\rangle. 
\end{equation}
Arguing along a similar line of logic as before, we identify the Pauli strings $P$ for which $\langle Pi, Pi, Pj, Pj|jjii\rangle \neq 0$. If this overlap is nonzero, then $P$ must be a string consisting solely of $X$ and $Y$ operators on the qubits for which $|i\rangle$ and $|j\rangle$ differ, and solely $I$ and $Z$ operators on the qubits for which $|i\rangle$ and $|j\rangle$ are the same. Furthermore, any such Pauli string satisfies $\langle Pi, Pi, Pj, Pj|jjii\rangle = 1$. One may see this by realizing that $P|i\rangle = e^{i \phi}|j\rangle$ for any such $P$, where $\phi\in \{0, \pi/2, \pi, 3\pi/2\}$. This implies that $P|j\rangle = e^{-i\phi}|i\rangle$, which gives $P|i\rangle P|i\rangle P|j\rangle P|j\rangle = e^{2i\phi} e^{-2i\phi} |jjii\rangle = |jjii\rangle$. Note that even though the Pauli strings in consideration may consist of all 4 Pauli operators, there are still only $2^{N_B}$ such strings because each qubit has a choice of only 2 Pauli operators. Therefore, there are $2^{N_B}$ pauli strings $P\in \mathcal{P}_{N_B}$ for which  $\langle Pi, Pi, Pj, Pj|jjii\rangle = 1$, and the overlap is zero for all others. We conclude that~\eqref{eq:flipping-perms-mat-el} becomes 
\begin{equation}
    \langle iijj|Q_BT_\rho^B|iijj\rangle = \frac{1}{d_B^2}\cdot 2^{N_B} = \frac{1}{d_B}
\end{equation}
if $\rho \in \{(13)(24), (14)(23), (1324), (1423)\}$. 

To summarize, there are exactly 8 permutations $\rho$, namely $\{\text{Id}, (12), (34), (12)(34), (13)(24), (14)(23), (1324), (1423)\}$, for which $\langle iijj|Q_BT_\rho^B|iijj\rangle = 1/d_B$, while $\langle iijj|Q_BT_\rho^B|iijj\rangle = 0$ for all other $\rho$. Furthermore, note that these permutations are the subgroup $\langle (1324), (12)\rangle$ of $S_4$.

We now revisit the first term from the numerator~\eqref{eq:num-former-term}. We compute that  
\begin{equation}\label{eq:former-term-heavy-lifting-calc}
\begin{split}
    \text{Tr}[S_A([P_i^{\otimes 2}\otimes P_j^{\otimes 2}]Q\Pi_4^{\rm sym}[P_i^{\otimes 2}\otimes P_j^{\otimes 2}])] &= \sum_{\rho \in \langle (1324), (12)\rangle} \frac{\text{Tr}[S_AQ_AT_\rho^A]}{d_B} \\
    &= \sum_{\rho \in \langle (1324), (12)\rangle} \frac{\text{Tr}[Q_AS_AT_\rho^A]}{d_B} \\
    &= \sum_{\rho \in \langle (1324), (12)\rangle} \frac{\text{Tr}[Q_AT_{(13)(24)}^AT_\rho^A]}{d_B} \\
    &= \sum_{\rho \in \langle (1324), (12)\rangle}  \frac{\text{Tr}[Q_A T_\rho^A]}{d_B} \\
    &= \frac{1}{d_A^2} \sum_{P\in\mathcal{P}_{N_A}} \sum_{\rho\in\langle (1324), (12)\rangle} \frac{\text{Tr}[P^{\otimes 4} T_\rho^A]}{d_B} \\
    &= \frac{1}{d_A^2} \sum_{P\in\mathcal{P}_{N_A}} \sum_{\rho\in\langle (1324), (12)\rangle} \frac{1}{d_B} \prod_{\alpha\in \text{cyc}(\rho)} \text{Tr}[P^{|\alpha|}] \\
    &= \frac{1}{d_A^2 d_B} \sum_{P\in\mathcal{P}_{N_A}} (\text{Tr}[P]^4 + 2\text{Tr}[P^2]\text{Tr}[P]^2 \\
    &\ \ \ \ \ \ \ \ \ \ \ \ + 3\text{Tr}[P^2]^2 + 2\text{Tr}[P^4]) \\
    &= \frac{1}{d_A^2d_B}(d_A^4 + 2d_A^3  + \sum_{P\in \mathcal{P}_{N_A}} (3 d_A^2 + 2d_A)) \\
    &= \frac{1}{d_B}(d_A^2 + 2d_A + 3d_A^2 + 2d_A) \\
    &= \frac{4d_A(d_A + 1)}{d_B}.
\end{split}
\end{equation}
The fourth equality is due to the fact that left multiplication by a group element determines a bijection from the group to itself. In the sixth equality, we define $\text{cyc}(\rho)$ as the unique set of disjoint cycles into which the permutation $\rho$ decomposes. We also use $|\alpha|$ to indicate the length of a cycle $\alpha$. The sixth equality then follows from the general fact that $\text{Tr}[O^{\otimes k} T_\rho] = \prod_{\alpha\in\text{cyc}(\rho)} \text{Tr}[O^{|\alpha|}]$ for any operator $O$ and $\rho\in S_k$. The eighth equality is derived from the observation that when $t$ is odd, we have $\text{Tr}[P^t] = 2^{N_A} = d_A$ for the Pauli string $P = I^{\otimes N_A}$, while $\text{Tr}[P^t] = 0$ for all other Pauli strings. On the other hand, when $t$ is even $P^t=I^{\otimes N_A}$, so $\text{Tr}[P^t] = 2^{N_A} = d_A$ in this case. Putting these identities together, one obtains the eighth equality.

Now assume that $i = j$. Following the work of~\eqref{eq:num-former-term}, we arrive at
\begin{equation}\label{eq:final-calc-from-num}
\begin{split}
    \text{Tr}[S_A([P_i^{\otimes 2}\otimes P_j^{\otimes 2}]Q\Pi_4^{\rm sym}[P_i^{\otimes 2}\otimes P_j^{\otimes 2}])] &= \sum_{\rho\in S_4} \text{Tr}[S_AQ_AT_\rho^A] \langle iiii|Q_B T_\rho^B|iiii\rangle \\
    &= \sum_{\rho\in S_4} \text{Tr}[S_AQ_AT_\rho^A] \langle iiii|Q_B |iiii\rangle \\
\end{split}
\end{equation}
As before, we expand the matrix element 
\begin{equation}
    \langle iiii|Q_B|iiii\rangle = \frac{1}{d_B^2} \sum_{P\in\mathcal{P}_{N_B}} \langle Pi,Pi,Pi,Pi|iiii\rangle. 
\end{equation}
We note that by an argument similar to a previous one, we have that $\langle Pi, Pi, Pi, Pi|iiii\rangle = 1$ for all Pauli strings $P\in \{I, Z\}^{\otimes N_B}$ and $\langle Pi, Pi, Pi, Pi|iiii\rangle = 0$ for all others. Since there are $d_B = 2^{N_B}$ strings in $\{I, Z\}^{\otimes N_B}$, it follows that $\langle iiii|Q_B|iiii\rangle = \frac{1}{d_B^2} d_B = \frac{1}{d_B}$. Going back to~\eqref{eq:final-calc-from-num}, and using a line of argumentation similar to that of~\eqref{eq:former-term-heavy-lifting-calc}, we obtain  
\begin{equation}\label{eq:i-eq-j-former-num-term}
\begin{split}
    \text{Tr}[S_A([P_i^{\otimes 2}\otimes P_j^{\otimes 2}]Q\Pi_4^{\rm sym}[P_i^{\otimes 2}\otimes P_j^{\otimes 2}])] 
    &= \frac{1}{d_A^2} \sum_{P\in \mathcal{P}_{N_A}} \sum_{\rho\in S_4} \frac{1}{d_B} \prod_{\alpha\in\text{cyc}(\rho)} \text{Tr}[P^{|\alpha|}] \\
    &= \frac{1}{d_A^2 d_B} \sum_{P\in \mathcal{P}_{N_A}} (\text{Tr}[P]^4 + 6\text{Tr}[P^2]\text{Tr}[P]^2 \\
    &\ \ \ \ \ \ \ \ \ \ \ \ + 3\text{Tr}[P^2]^2 + 8\text{Tr}[P^3]\text{Tr}[P] + 6\text{Tr}[P^4]) \\
    &= \frac{1}{d_A^2 d_B} (d_A^4 + 6 d_A^3 + 8d_A^2 + \sum_{P\in \mathcal{P}_{N_A}} (3d_A^2 + 6d_A) \\
    &= \frac{1}{d_B}(d_A^2 + 6d_A + 8 + 3d_A^2 + 6d_A) \\
    &= \frac{4(d_A + 2)(d_A + 1)}{d_B}.
\end{split}
\end{equation}
The second equality follows from counting the elements of each possible cycle type in $S_4$.

Finally, putting~\eqref{eq:num-second-term-final-expression},~\eqref{eq:former-term-heavy-lifting-calc}, and~\eqref{eq:i-eq-j-former-num-term} together, we get that that the numerator expression evaluates to
\begin{equation}\label{eq:num-term-final-expression}
    \text{Tr}[S_A([P_i^{\otimes 2}\otimes P_j^{\otimes 2}]\langle C^{\otimes 4}\psi^{\otimes 4} C^{\dagger \otimes 4}\rangle_{\mathcal{C}_N} [P_i^{\otimes 2}\otimes P_j^{\otimes 2}])] = \begin{cases}
        d_A(d_A + 1)\bigg(\frac{4a}{d_B} + 2b\bigg), & i\neq j \\
        (d_A + 1)(d_A + 2)\bigg(\frac{4a}{d_B} + bd_A(d_A + 3)\bigg), & i = j
    \end{cases}
\end{equation}

\subsubsection{Final Computations for Term I} 

Using the results from the previous three sections, term I now reads
\begin{equation}
    \begin{split}
        (\textnormal{I}) &= \sum_{\stackrel{i,j \in \{0,1\}^{N_B}}{i\neq j}} \frac{ \text{Tr}[S_A([P_i^{\otimes 2}\otimes P_j^{\otimes 2}]\langle C^{\otimes 4}\psi^{\otimes 4} C^{\dagger \otimes 4}\rangle_{\mathcal{C}_N}[P_i^{\otimes 2}\otimes P_j^{\otimes 2}])]}
    {\text{Tr}[(P_i\otimes P_j)\langle C^{\otimes 2} \psi^{\otimes 2} C^{\dagger \otimes 2}\rangle_{\mathcal{C}_N}]} \\
    &\ \ \ \ + \sum_{\stackrel{i,j \in \{0,1\}^{N_B}}{i=j}} \frac{ \text{Tr}[S_A([P_i^{\otimes 2}\otimes P_j^{\otimes 2}]\langle C^{\otimes 4}\psi^{\otimes 4} C^{\dagger \otimes 4}\rangle_{\mathcal{C}_N}[P_i^{\otimes 2}\otimes P_j^{\otimes 2}])]}
    {\text{Tr}[(P_i\otimes P_j)\langle C^{\otimes 2} \psi^{\otimes 2} C^{\dagger \otimes 2}\rangle_{\mathcal{C}_N}]} \\
    &= \sum_{\stackrel{i,j \in \{0,1\}^{N_B}}{i\neq j}} \frac{d_A(d_A + 1)(4a/d_B + 2b)}{\frac{d_A}{d_B(d_Ad_B+1)}} + 
    \sum_{\stackrel{i,j \in \{0,1\}^{N_B}}{i= j}} \frac{(d_A + 1)(d_A + 2)(4a/d_B + bd_A(d_A + 3))}{\frac{d_A + 1}{d_B(d_Ad_B + 1)}} \\
    &= d_B^2(d_B - 1)(d_A d_B + 1)(d_A + 1)(4a/d_B + 2b) + d_B^2(d_Ad_B + 1)(d_A + 2)(4a/d_B + bd_A(d_A + 3))
    \end{split}
\end{equation}

\subsubsection{Mean Quotient Approximation for Term I}\label{mean-quotient-approx}
We now give a justification for our approximation of the quotient expressions in term (I). For each $C\in \mathcal{C}_N$, define 
\begin{align}
    x_C &\equiv \text{Tr}[S_A([P_i^{\otimes 2}\otimes P_j^{\otimes 2}][C^{\otimes 4}\psi^{\otimes 4} C^{\dagger \otimes 4}][P_i^{\otimes 2}\otimes P_j^{\otimes 2}])] \\
    y_C &\equiv \text{Tr}[(P_i\otimes P_j)(C^{\otimes 2}\psi^{\otimes 2} C^{\dagger \otimes 2})].
\end{align}
We note that $\langle x_C/y_C \rangle_{\mathcal{C}_N} \approx \langle x_C\rangle_{\mathcal{C}_N}/\langle y_C\rangle_{\mathcal{C}_N}$ if the standard deviation $\sigma_y = \sqrt{\langle y_C^2\rangle_{\mathcal{C}_N}- \langle y_C\rangle_{\mathcal{C}_N}^2}$ satisfies $\sigma_y \ll \langle y_C \rangle_{\mathcal{C}_N}$. Thus, if we can show that $\sigma_y \ll \langle y_C \rangle_{\mathcal{C}_N}$, the approximation used in the fourth line of \eqref{eq:Tr-rho2-calc} would be justified.

To this end, we now compute the ratio $r\equiv \sigma_y^2/\langle y_C\rangle_{\mathcal{C}_N}^2$. The mean $\langle y_C\rangle_{\mathcal{C}_N}$ is given by~\eqref{eq:term1-denom}. Observe that 
\begin{equation}\label{eq:SMom}
    \begin{split}
        \langle y_C^2 \rangle_{\mathcal{C}_N} &=
        \langle \text{Tr}[(P_i \otimes P_j)C^{\otimes 2} \psi^{\otimes 2} C^{\dagger \otimes 2}]^2 \rangle_{\mathcal{C}_N}\\ &=
        \langle \text{Tr}[(P_i^{\otimes 2} \otimes P_j^{\otimes 2}) C^{\otimes 4} \psi^{\otimes 4} C^{\dagger \otimes 4}] \rangle_{\mathcal{C}_N} \\
        &=
        \text{Tr}[(P_i^{\otimes 2}\otimes P_j^{\otimes 2})\langle C^{\otimes 4} \psi^{\otimes 4} C^{\dagger \otimes 4}\rangle_{\mathcal{C}_N}]  \\
        &= a \text{Tr}[(P_i^{\otimes 2} \otimes P_j^{\otimes 2})Q\Pi_4^{\rm sym}] \\
        &\ \ \ \ \ \ \ \ + b\text{Tr}[(P_i^{\otimes 2} \otimes P_j^{\otimes 2}) \Pi_4^{\rm sym}].
    \end{split}
\end{equation}

First, we assume $i\neq j$. Using the usual arguments outlined earlier, we have
\begin{equation}\label{eq:Pi-term}
    \begin{split}
        \text{Tr}[(P_i^{\otimes 2} \otimes P_j^{\otimes 2}) \Pi_4^{\rm sym}] &= 
        \sum_{\rho\in S_4} \text{Tr}[T_\rho^A \otimes |iijj\rangle\langle iijj|T_\rho^B] \\
        &= \sum_{\rho\in S_4} \text{Tr}[T_\rho^A] \langle iijj|T_\rho^B|iijj\rangle \\
        &= \sum_{\rho\in \langle (12), (34)\rangle} \text{Tr}[T_\rho^A] \\
        &= \sum_{\rho, \sigma\in S_2} \text{Tr}[T_\rho^{A, 12} \otimes T_\sigma^{A, 34}] \\
        &= \text{Tr}[\Pi_{2,A}^{sym} \otimes \Pi_{2,A}^{sym}] \\
        &= d_A^2(d_A + 1)^2.
    \end{split}
\end{equation}
We also have
\begin{equation}\label{eq:QPi-term}
    \begin{split}
        \text{Tr}[(P_i^{\otimes 2} \otimes P_j^{\otimes 2})Q\Pi_4^{\rm sym}] &=
        \sum_{\rho\in S_4} \text{Tr}[Q_AT_\rho^A] \langle iijj|Q_BT_\rho^B|iijj\rangle \\
        &=
        \sum_{\rho\in \langle (1324), (12)\rangle} \frac{\text{Tr}[Q_A T_\rho^B]}{d_B} \\
        &= \frac{4d_A(d_A + 1)}{d_B}.
    \end{split}
\end{equation}
In the above computation, the quantity on the RHS of the second line is identical to one that appears in~\eqref{eq:former-term-heavy-lifting-calc}, so we may simply use the result of that calculation. Therefore, using~\eqref{eq:Pi-term} and~\eqref{eq:QPi-term}, equation~\eqref{eq:SMom} becomes
\begin{equation}
    \langle y_C^2 \rangle_{\mathcal{C}_N} = \frac{4ad_A(d_A + 1)}{d_B } + bd_A^2(d_A + 1)^2.
\end{equation}
Using the definition~\eqref{eq:alpha-beta-defs} of $a$ and $b$ together with the fact that $\|\Xi(|\psi\rangle)\|_2^2 = [1-M_{\rm lin}(|\psi\rangle)]/d$, one then gets that
\begin{equation}
    \begin{split}
        \langle y_C^2 \rangle_{\mathcal{C}_N}
        &= \frac{4d_A(d_A + 1)}{d_B} \frac{\|\Xi(|\psi\rangle)\|_2^2}{4(d+1)(d+2)} -
        \frac{4d_A(d_A + 1)}{d_B} \frac{1-\|\Xi(|\psi\rangle)\|_2^2}{(d^2 - 1)(d+2)(d+4)} \\
        &\ \ \ \ +d_A^2(d_A + 1)^2 \frac{1-\|\Xi(|\psi\rangle)\|_2^2}{(d^2 - 1)(d+2)(d+4)} \\
        &= \frac{d_A(d_A + 1)(1- M_{\rm lin}(|\psi\rangle)}{d_B d(d+1)(d+2)} -
        \frac{4d_A(d_A + 1)[d- 1 + M_{\rm lin}(|\psi\rangle)]}{d_B d(d^2 - 1)(d+2)(d+4)} \\
        &\ \ \ \ + \frac{d_A^2(d_A + 1)^2[d - 1 + M_{\rm lin}(|\psi\rangle)]}{d(d^2 - 1)(d+2)(d+4)} \\
        &= \frac{1}{d_B^4} + \Theta\bigg(\frac{1}{d_Ad_B^4} \bigg) 
    \end{split}
\end{equation}
where the last equality relies on the fact that $0\leq M_{\rm lin}(|\phi\rangle)< 1$ for all states $|\phi\rangle$.  
Hence, the ratio $r$ of the variance to the square of the mean of $y_C$ is 
\begin{equation}
    \begin{split}
        r &= 
        \frac{ \langle y_C^2\rangle_{\mathcal{C}_N} - \langle y_C \rangle_{\mathcal{C}_N}^2}{\langle y_C \rangle_{\mathcal{C}_N}^2} \\
        &= \frac{\bigg[\frac{1}{d_B^4} + \Theta\bigg(\frac{1}{d_Ad_B^4} \bigg) \bigg] - \bigg[\frac{1}{d_B^4} + \Theta\bigg(\frac{1}{d_A d_B^5}\bigg)\bigg]}{\frac{1}{d_B^4} + \Theta\bigg(\frac{1}{d_A d_B^5}\bigg)} \\
        &= \Theta\bigg(\frac{1}{d_A}\bigg)
    \end{split}
\end{equation}
Hence, for a fixed value of $M_{\rm lin}$, we have that $r\to 0$ as $d_A\to\infty$. By re-purposing prior calculations, one can also show that $r$ vanishes as $d_A$ grows large in the $i = j$ case as well.  This shows that our approximation holds for large $d_A$. 

\subsection{Evaluation of Term (II)}

We now wish to evaluate
\begin{equation}\label{eq:lin-term}
\begin{split}
    \langle \text{Tr}[\rho_C^{(2)}\rho_H^{(2)}]\rangle_{\mathcal{C}_N} &= 
    \bigg\langle\sum_{i\in\{0,1\}^{N_B}} \text{Tr}\bigg\{ \frac{\text{Tr}_B[P_iC\psi C^\dagger P_i]^{\otimes 2}}{\text{Tr}[P_iC\psi C^\dagger]} \rho_H^{(2)}\bigg\}\bigg\rangle_{\mathcal{C}_N} \\
    &=
    \sum_{i\in\{0,1\}^{N_B}} \bigg\langle\frac{\text{Tr}[S_A ([P_i^{\otimes 2} C^{\otimes 2} \psi^{\otimes 2} C^{\dagger \otimes 2} P_i^{\otimes 2}] \otimes [\rho_H^{(2)}\otimes (I_B/d_B)])]}{\text{Tr}[P_i C\psi C^\dagger]}\bigg\rangle_{\mathcal{C}_N} \\
    &\approx 
    \sum_{i\in\{0,1\}^{N_B}} \frac{\langle \text{Tr}[S_A([P_i^{\otimes 2} C^{\otimes 2} \psi^{\otimes 2} C^{\dagger \otimes 2} P_i^{\otimes 2}] \otimes [\rho_H^{(2)}\otimes (I_B/d_B)])] \rangle_{\mathcal{C}_N}}{\langle \text{Tr}[P_i C\psi C^\dagger]\rangle_{\mathcal{C}_N}} \\
    &=
    \sum_{i\in\{0,1\}^{N_B}} \frac{ \text{Tr}[S_A([P_i^{\otimes 2}\langle C^{\otimes 2} \psi^{\otimes 2} C^{\dagger \otimes 2} \rangle_{\mathcal{C}_N} P_i^{\otimes 2}]\otimes [\rho_H^{(2)}\otimes (I_B/d_B)])]}{\text{Tr}[P_i \langle C\psi C^\dagger\rangle_{\mathcal{C}_N}]}.
\end{split}
\end{equation}
As before, to show that the approximation on the third line of~\eqref{eq:lin-term} holds, we compute the standard deviation of $y_C \equiv \text{Tr}[P_iC\psi C^\dagger]$. To do so, we first calculate $\langle y_C^2\rangle_{\mathcal{C}_N}$:
\begin{equation}
    \begin{split}
        \langle  \text{Tr}[P_iC\psi C^\dagger]^2 \rangle_{\mathcal{C}_N} &=
        \langle  \text{Tr}[P_i^{\otimes 2} C^{\otimes 2}\psi^{\otimes 2} C^{\otimes 2 \dagger}] \rangle\\ &=
        \text{Tr}[P_i^{\otimes 2} \langle C^{\otimes 2}\psi^{\otimes 2} C^{\otimes 2 \dagger}\rangle] \\ &=
        \sum_{\rho\in S_2} \frac{\text{Tr}[P_i^{\otimes 2} T_\rho]}{d_Ad_B(d_Ad_B + 1)} \\ &=
        \sum_{\rho\in S_2} \frac{\text{Tr}[(I_A \otimes |i\rangle\langle i|_B)^{\otimes 2} T_\rho^A \otimes T_\rho^B]}{d_Ad_B(d_Ad_B +1)} \\ &=
        \sum_{\rho\in S_2} \frac{\text{Tr}[T_\rho^A \otimes |ii\rangle\langle ii|_B T_\rho^B]}{d_Ad_B(d_Ad_B + 1)} \\ &=
        \sum_{\rho\in S_2} \frac{\text{Tr}[T_\rho^A]}{d_Ad_B(d_Ad_B + 1)} \\ &=
        \frac{d_A(d_A + 1)}{d_Ad_B(d_Ad_B + 1)} \\ &=
        \frac{d_A + 1}{d_B(d_Ad_B + 1)}.
    \end{split}
\end{equation} Next, we compute $\langle y_C\rangle_{\mathcal{C}_N}$:
\begin{equation}\label{eq:lin-term-num}
    \begin{split}
        \langle \text{Tr}[P_i  C\psi C^\dagger]\rangle_{\mathcal{C}_N} &= \text{Tr}[P_i\langle C\psi C^\dagger\rangle_{\mathcal{C}_N}] \\ &= \text{Tr}\bigg[P_i \frac{I_{\mathcal{H}_A\otimes\mathcal{H}_B}}{d_Ad_B}\bigg] \\
        &= \text{Tr}\bigg[\frac{I_A \otimes |i\rangle\langle i|_B}{d_Ad_B}\bigg] \\
        &= \frac{1}{d_B}.
    \end{split}
\end{equation}
It follows that
\begin{equation}
    \sigma_y^2 = \langle y_C^2\rangle_{\mathcal{C}_N} - \langle y_C\rangle^2_{\mathcal{C}_N} = \frac{d_A + 1}{d_B(d_Ad_B + 1)} - \frac{1}{d_B^2} = \Theta\bigg(\frac{1}{d_Ad_B^2}\bigg).
\end{equation}
Hence, $\sigma_y^2/\langle y_C \rangle^2_{\mathcal{C}_N} \to 0$ as $d_A \to \infty$, and we can safely conclude that our approximation holds in this limit.
Next, we calculate the numerator:
\begin{equation}\label{eq:lin-term-den}
    \begin{split}
        \text{Tr}[S_A([P_i^{\otimes 2}\langle C^{\otimes 2} \psi^{\otimes 2} C^{\dagger \otimes 2} \rangle_{\mathcal{C}_N} &P_i^{\otimes 2}] \otimes [\rho_H^{(2)}\otimes (I_B/d_B)])] \\
        &= \sum_{\rho\in S_2}
        \frac{\text{Tr}[S_A([P_i^{\otimes 2} T_\rho P_i^{\otimes 2}]\otimes [\rho_H^{(2)}\otimes (I_B/d_B)])]}{d_Ad_B(d_Ad_B + 1)} \\
        &=
        \sum_{\rho\in S_2}
        \frac{\text{Tr}[S_A([(I_A\otimes |i\rangle\langle i|_B)^{\otimes 2} (T_\rho^A \otimes T_\rho^B) (I_A \otimes |i\rangle\langle i|_B)^{\otimes 2}]\otimes [\rho_H^{(2)}\otimes (I_B/d_B)])]}{d_Ad_B(d_Ad_B + 1)} \\
        &=
        \sum_{\rho\in S_2}
        \frac{\text{Tr}[(S_A[T_\rho^A\otimes \rho_H^{(2)}])\otimes (|ii\rangle\langle ii|_BT_\rho^B|ii\rangle\langle ii|_B\otimes [I_B/d_B])]}{d_Ad_B(d_Ad_B + 1)} \\
        &=
        \sum_{\rho\in S_2}
        \frac{\text{Tr}[S_A(T_\rho^A\otimes \rho_H^{(2)})]\langle ii|_BT_\rho^B|ii\rangle_B}{d_Ad_B(d_Ad_B + 1)} \\
        &=
        \sum_{\rho\in S_2}
        \frac{\text{Tr}[S_A(T_\rho^A\otimes \rho_H^{(2)})]}{d_Ad_B(d_Ad_B + 1)} \\
        &=
        \sum_{\rho\in S_2}
        \frac{\text{Tr}[T_\rho^A \rho_H^{(2)}]}{d_Ad_B(d_Ad_B + 1)} \\
        &= 
        \sum_{\rho\in S_2}
        \frac{\text{Tr}[\rho_H^{(2)}]}{d_Ad_B(d_Ad_B + 1)} \\
        &= \frac{2}{d_Ad_B(d_Ad_B + 1)}.
    \end{split}
\end{equation}
Putting~\eqref{eq:lin-term},~\eqref{eq:lin-term-den}, and~\eqref{eq:lin-term-num} together, we get
\begin{equation}
    \langle \text{Tr}[\rho_C^{(2)}\rho_H^{(2)}]\rangle_{\mathcal{C}_N} = \sum_{i\in \{0,1\}^{N_B}} \frac{\frac{2}{d_Ad_B(d_Ad_B + 1)}}{\frac{1}{d_B}} 
    = \frac{2d_B}{d_A(d_Ad_B + 1)}
\end{equation}

\subsection{Evaluation of Term (III)}

Term (III) evaluates simply to 
\begin{equation}\label{eq:Tr-RhoH}
    \begin{split}
        \text{Tr}[(\rho_H^{(2)})^2] &=
        \frac{\text{Tr}[(\Pi_{2, A}^{sym})^2]}{\text{Tr}[\Pi_{2, A}^{sym}]^2} =
        \frac{\text{Tr}[2\Pi_{2, A}^{sym}]}{\text{Tr}[\Pi_{2, A}^{sym}]^2} =
        \frac{2}{\text{Tr}[\Pi_{2, A}^{sym}]}=
        \frac{2}{d_A(d_A + 1)}.
    \end{split}
\end{equation}

\subsection{Writing Average Hilbert-Schmidt Distance in Terms of Magic}
Now that we have calculated terms I, II, and III, we can write $\langle \|\rho_C^{(2)} - \rho_H^{(2)}\|_2^2\rangle_{\mathcal{C}_N} = (\text{I}) - 2(\text{II}) + (\text{III})$ in terms of $M_{\rm lin}(|\psi\rangle) = 1 - d_Ad_B\|\Xi(|\psi\rangle)\|_2^2$. Applying some straightforward but tedious algebra to the above results, we get that
\begin{equation}\label{eq:x-y-eq-average-2-designness}
    \langle [d_{\rm HS}^{(2)}(\mathcal{E}_{C|\psi\rangle})]^2 \rangle_{\mathcal{C}_N} = 
    \langle \|\rho_C^{(2)} - \rho_H^{(2)}\|_2^2\rangle_{\mathcal{C}_N} = x + y\|\Xi(|\psi\rangle)\|_2^2,
\end{equation}
where 
\begin{equation}
\begin{split}
    x &= \frac{2d_B(d_B - 1)(d_B - 2)(d_A + 1)}{(d-1)(d+2)(d+4)} +
    \frac{d_B(d_A + 2)[d(d_A + 3) - 4]}{(d-1)(d+2)(d+4)} - \frac{4d_B^2}{d(d+1)} + \frac{2}{d_A(d_A + 1)}.
\end{split}
\end{equation}
and
\begin{equation}
    \begin{split}
        y &= -
        \frac{2d_B(d_B - 1)(d_A + 1)(d_B - 2)}{(d-1)(d+2)(d+4)} -
        \frac{d_B(d_A + 2)[d(d_A  + 3) - 4]}{(d-1)(d+2)(d+4)}
        +\frac{d_B[(d_B - 1)(d_A + 1) + d_A + 2]}{d+2} .
    \end{split}
\end{equation}
Note that~\eqref{eq:x-y-eq-average-2-designness} may be re-written as
\begin{equation}
    \langle \|\rho^{(2)} - \rho_H^{(2)}\|_2^2\rangle_{\mathcal{C}_N} = \alpha - \beta M_{\rm lin}(|\psi\rangle) , 
\end{equation}
where
\begin{equation}\label{eq:alphaLin}
    \begin{split}
        \alpha &= x + \frac{y}{d} \\
        &=\frac{2d_B(d_B - 1)(d_B - 2)(d_A + 1)}{d(d+2)(d+4)} 
        + \frac{d_B(d_A + 2)[d(d_A + 3) - 4]}{d(d+2)(d+4)}
        + \frac{d_B(d + d_B + 1)}{d(d+ 2)} - \frac{4d_B^2}{d(d+1)} + \frac{2}{d_A(d_A + 1)}.
    \end{split}
\end{equation}
and
\begin{equation}\label{eq:betaLin}
    \begin{split}
        \beta &= \frac{y}{d}\\
        &= \frac{d_B(d + d_B + 1)}{d(d+2)} - \frac{2d_B(d_B - 1)(d_B - 2)(d_A + 1)}{d(d- 1)(d+2)(d+4)}
        -
        \frac{d_B(d_A + 2)[d(d_A + 3) - 4]}{d(d-1)(d+2)(d+4)}.
    \end{split}
\end{equation}
Observe that the second and third terms of~\eqref{eq:alphaLin} are the dominant terms, scaling as $1/d_B$ and $1/d_A$, respectively. The first term of~\eqref{eq:betaLin} is dominant and identical to the third of~\eqref{eq:alphaLin}, scaling as $1/d_A$.

\subsection{Estimate of Total Error}
Define 
\begin{align}
        x_C^{ij} &\equiv
     \text{Tr}[S_A([P_i^{\otimes 2}\otimes P_j^{\otimes 2}][C^{\otimes 4}\psi^{\otimes 4} C^{\dagger \otimes 4}][P_i^{\otimes 2}\otimes P_j^{\otimes 2}])] \\
     y_C^{ij} &\equiv \text{Tr}[(P_i\otimes P_j)(C^{\otimes 2}\psi^{\otimes 2} C^{\dagger \otimes 2})]
\end{align}
We now compute the errors associated with our approximation of the summands in term (I). The error associated with the $ij$-th cross term in~\eqref{eq:Tr-rho2-calc} is
\begin{equation}\label{eq:cross-term-error-calc}
    \begin{split}
        \varepsilon_{ij} &\equiv \bigg\langle \frac{x^{ij}_C}{y^{ij}_C}\bigg\rangle_{\mathcal{C}_N} -
        \frac{\langle x_{C}^{ij}\rangle_{\mathcal{C}_N}}{\langle y_{C}^{ij}\rangle_{\mathcal{C}_N}} \\
        &= \bigg\langle
        x_C^{ij} \bigg(\frac{1}{y_C^{ij}} - \frac{1}{\langle y_C^{ij}\rangle_{\mathcal{C}_N}}\bigg)\bigg
        \rangle_{\mathcal{C}_N} \\
        &= \bigg \langle \frac{x_C^{ij}}{(y_C^{ij})^2} 
        \bigg(y_C^{ij} - \frac{(y_C^{ij})^2}{\langle y_C^{ij}\rangle_{\mathcal{C}_N}}\bigg)\bigg
        \rangle_{\mathcal{C}_N} \\
        &\approx \bigg\langle \frac{x_C^{ij}}{(y_C^{ij})^2} \bigg\rangle_{\mathcal{C}_N} 
        \bigg \langle y_C^{ij} - \frac{(y_C^{ij})^2}{\langle y_C^{ij}\rangle_{\mathcal{C}_N}} \bigg\rangle_{\mathcal{C_N}} \\
        &\approx \bigg \langle \frac{x_C^{ij}}{y_C^{ij}}\bigg\rangle_{\mathcal{C}_N} \frac{1}{\langle y_C^{ij}\rangle_{\mathcal{C}_N}}
        \bigg \langle y_C^{ij} - \frac{(y_C^{ij})^2}{\langle y_C^{ij}\rangle_{\mathcal{C}_N}} \bigg\rangle_{\mathcal{C_N}} \\ 
        &= -\bigg\langle \frac{x_C^{ij}}{y_C^{ij}}\bigg\rangle_{\mathcal{C}_N}\frac{\langle (y_C^{ij})^2\rangle_{\mathcal{C}_N} - \langle y_C^{ij} \rangle_{\mathcal{C}_N}^2}{ \langle y_C^{ij}\rangle_{\mathcal{C}_N}^2} \\
        &= -\bigg\langle \frac{x_C^{ij}}{y_C^{ij}}\bigg\rangle_{\mathcal{C}_N} \frac{\sigma_{y_C^{ij}}^2}{\langle y_C^{ij} \rangle_{\mathcal{C}_N}^2} \\
        &= -\bigg\langle \frac{x_C^{ij}}{y_C^{ij}}\bigg\rangle_{\mathcal{C}_N} \Theta\bigg(\frac{1}{d_A}\bigg),
    \end{split}
\end{equation}
where it is understood that $\Theta(1/d_A)$ is taken to be a positive function. 
The fourth and fifth lines may be derived as follows from Lemma 4 of~\cite{Cotler2023}, though we do not prove the details formally. The basic idea of this lemma is that the normalized state obtained by projectively measuring out part of a Haar-random state and the probability associated with this measurement outcome are independent random variables. The uniformly weighted Clifford group matches the moments of the Haar ensemble up to the third and approximates its fourth moment for large systems~\cite{Zhu2017,Tirrito2023}, so $x_C^{ij}/(y_C^{ij})^2$ and any function solely dependent on $y_C^{ij}$ should be approximately independent random variables. It follows that our expression for $\varepsilon_{ij}$ should factor as it does on the fouth line above, while the fifth line follows from the factorization $\langle x_C^{ij}/y_C^{ij}\rangle = \langle y_C^{ij} x_C^{ij}/(y_C^{ij})^2   \rangle \approx \langle y_C^{ij}\rangle \langle x_C^{ij}/(y_C^{ij})^2\rangle $. 

A similar computation shows that the approximation errors of the summands of term (II) in~\eqref{eq:lin-term} are also proportional to the their respective exact values by a factor of $-\Theta(1/d_A)$. If our approximations for terms (I) and (II) are reasonable, we can expect that the order of the exact term (III) in $1/d_A$ is at least that of terms (I) and (II). It follows that the total approximation error, which we define to be $\varepsilon \equiv \langle [d_{\rm HS}^{(2)}(\mathcal{E}_{C|\psi\rangle})]^2 \rangle_{\mathcal{C}_N} - [\alpha(d_A, d_B) - \beta(d_A, d_B) M_{\rm lin}(|\psi\rangle)]$, satisfies
\begin{equation}\label{eq:total-error-eq}
    \frac{\varepsilon}{\langle [d_{\rm HS}^{(2)}(\mathcal{E}_{C|\psi\rangle})]^2 \rangle_{\mathcal{C}_N}} =  \frac{-\Theta\bigg(\frac{1}{d_A}\bigg)\times[\text{(I)} - 2\text{(II)}]}{\text{(I)} - 2\text{(II)} + \text{(III)}} 
    = -\Theta\bigg(\frac{1}{d_A}\bigg).
\end{equation}
Since we can also expect that term (II) is of greater order in $1/d_A$ than term (I), we may take $\text{(I)} - 2\text{(II)}$ to be positive. This implies that the total approximation error is negative, as indicated in \eqref{eq:total-error-eq}.

Finally, we note that Fig.~\ref{fig:linear} is consistent with the above observations. Indeed, the simulated values of $\langle [d_{\rm HS}^{(2)}(\mathcal{E}_{C|\psi\rangle})]^2 \rangle_{\mathcal{C}_N}$ consistently lie below those of our linear approximation, which supports the conclusion that $\varepsilon$ is negative. Furthermore, the gap between these values decreases as $\langle [d_{\rm HS}^{(2)}(\mathcal{E}_{C|\psi\rangle})]^2 \rangle_{\mathcal{C}_N}$ decreases, which is consistent with the relationship between the two that is implied by \eqref{eq:total-error-eq}. Finally, \eqref{eq:total-error-eq} gives the right order of magnitude for the deviation. It should be noted that in the $i=j$ case, the ratio $\sigma^2_{y^{ij}}/\langle y_C^{ij} \rangle^2_{\mathcal{C}_N}$ in \eqref{eq:cross-term-error-calc} can be estimated more accurately by a certain multiple of $1/d_A$. By then numerically adding up all approximation errors as determined by \eqref{eq:cross-term-error-calc}, we are able to account for the gap more precisely, though we cannot mathematically express this situation as compactly as we can with the expression \eqref{eq:total-error-eq}.
Indeed, we account for this gap with Fig.~\ref{fig:linear} by plotting via the light dashed line the analytic approximation minus the error estimated in this way.

\subsection{Concentration of Measure}

\begin{figure}
    \centering
    \includegraphics[width=0.5\linewidth]{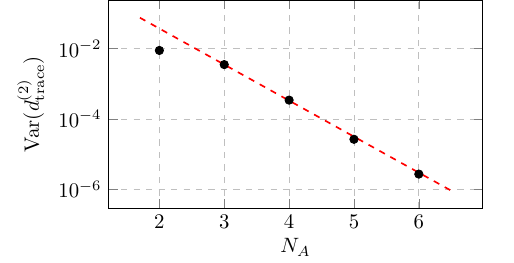}
    \caption{
    We plot the variance of the trace distance between second moments of the Haar ensemble and the projected ensemble of states obtained by randomly sampling Clifford circuits of fixed depth $L = 100$. The procedure for generating the circuits is described in the main text. $N_B$ is fixed at $4$. For $N_A = 2,3,4,5$, the variance is obtained from a sample consisting of 1000 circuits. For $N_A = 6$, the sample consists of 200 circuits in the interest of reducing runtime.
    }
    \label{fig:variance-na}
\end{figure}

Our analytical results concern the averaged \textit{square} of the distance between the second moments of the projected ensemble $\mathcal{E}_{C|\Psi\rangle}$ and the Haar ensemble. Due to concentration of measure, we expect that this average-of-squares is approximately equal to the square-of-averages for large system sizes, indirectly giving us a way of treating the averaged distance itself. Here, we provide numerical evidence for this conjecture.

Using the method described in the main text, we randomly construct Clifford circuits of a fixed depth $L = 100$ initialized to the product state $|\Psi(\theta)\rangle = 2^{-N/2}(|0\rangle + e^{i\theta}|1\rangle)^{\otimes N}$ for $\theta = \pi/4$, which maximizes magic for product states of this form. We compute the projected ensemble for each randomly sampled circuit. For varying sizes $N_A$ of the unmeasured system, we compute the variance of the normalized trace distance between the second moments of the projected ensemble and the Haar ensemble across a sample of random circuits. Note that the variance is exactly equal to the difference between the average-of-squares and square-of-averages for a sample. These variances are plotted in Fig.~\ref{fig:variance-na}, and show a clear exponential decay with respect to $N_A$ (size of the target random state ensemble subsystem). 

Here, we only compute the variance over circuits of \textit{fixed} depth to reduce runtime for larger $N_A$, while in the main text, we extrapolated the sample means to infinite depth to estimate the average over the Clifford group in Fig.~3 (of the main text). However, the decay in the variance already becomes clear at a circuit depth of $L = 100$. Thus, our numerical experiments suggest that if the size of the $A$ subsystem is sufficiently large, then the squared average of the distance from Haar is approximately equal to the averaged square of this distance.

\subsection{Clifford Rotations Without Measurements} 

As discussed in the main text, we introduce the projected ensemble into our work to suggest a method for efficient generation of approximate state designs via measurements and to establish the viewpoint that magic quantifies the internal correlations of a quantum system. However, it is natural to consider the randomness of the ensemble of all states $C|\psi\rangle$, where $C$ is a Clifford operator drawn uniformly at random from the Clifford group, \textit{without} measurements performed on the $B$ subsystem. While an understanding of how the randomness of this ensemble depends on the magic of $|\psi\rangle$ is not essential for our main results, we nonetheless include a treatment of this problem for the sake of completeness. In particular, we show here that the distance between the 4th moments of the Haar distribution and the ensemble of all Clifford rotations of a state $|\psi\rangle \in \mathcal{H}$, $\dim \mathcal{H} = d$, is given by
\begin{equation}
    \left \|\langle C^{\otimes 4} \psi C^{\dagger \otimes 4} \rangle_{\mathcal{C}_N} - \rho_H^{(4)} \right \|_2^2 \sim \frac{6(1-M_{\rm lin}(|\psi\rangle))^2}{d^4}
\end{equation}
to leading order in $1/d$.

Writing $\rho_C^{(4)} \coloneqq \langle C^{\otimes 4} \psi C^{\dagger \otimes 4}\rangle_{\mathcal{C}_N}$, we have
\begin{align}
    \left \|\langle C^{\otimes 4} \psi C^{\dagger \otimes 4} \rangle_{\mathcal{C}_N} - \rho_H^{(4)} \right \|_2^2 &= 
    \underbrace{{\rm Tr}[(\rho_C^{(4)})^2]}_{(i)} - 2\underbrace{{\rm Tr}[\rho_C^{(4)}\rho_H^{(4)}]}_{(ii)} + \underbrace{{\rm Tr}[(\rho_H^{(4)})^2]}_{(iii)} 
\end{align}
We calculate the above expression term-by-term as before. Recalling that $\rho_C^{(4)} = aQ\Pi_4^{\rm sym} + b\Pi_4^{\rm sym}$ from equation~\eqref{eq:clifford-4th-average}, we find 
\begin{align}
    (i) &\coloneqq  
    {\rm Tr}[(\rho_C^{(4)})^2] \\
    &= {\rm Tr}[(aQ\Pi_4^{\rm sym} + b \Pi_4^{\rm sym})^2] \\
    &= a^2 {\rm Tr}[Q\Pi_4^{\rm sym}Q\Pi_4^{\rm sym}] +
    2ab {\rm Tr}[Q(\Pi_4^{\rm sym})^2] + b^2{\rm Tr}[(\Pi_4^{\rm sym})^2] \\
    &= 24(a^2 {\rm Tr}[Q \Pi_4^{\rm sym}] + 2ab{\rm Tr}[Q\Pi_4^{\rm sym}] + b^2{\rm Tr}[\Pi_4^{\rm sym}]) \label{eq:projector-simplification},
\end{align}
where~\eqref{eq:projector-simplification} follows from the fact that $[Q, \Pi_4^{\rm sym}] = 0$, the projector relation $Q^2 = Q$, and $(\Pi_4^{\rm sym})^2 = 24\Pi_4^{\rm sym}$. 
The term ${\rm Tr}[Q\Pi_4^{\rm sym}]$ may be evaluated using calculations from our main results:
\begin{align}
    {\rm Tr}[Q\Pi_4^{\rm sym}] &= \frac{1}{d^2} \sum_{P \in \mathcal{P}_N} \sum_{\rho \in S^4} {\rm Tr}[P^{\otimes 4} T_\rho] \\
    &= \frac{1}{d^2} \sum_{P \in \mathcal{P}_N} \sum_{\rho \in S^4} \prod_{\alpha \in {\rm cyc}(\rho)} {\rm Tr}(P^{|\alpha|}) \\
    &= \frac{1}{d^2} \sum_{P \in \mathcal{P}_N}\left( {\rm Tr}[P]^4 + 6 {\rm Tr}[P^2]{\rm Tr}[P]^2 + 
    3{\rm Tr}[P^2]^2 + 8 {\rm Tr}[P^3]{\rm Tr}[P] + 6{\rm Tr}[P^4] \right)  \\
    &= \frac{1}{d^2}\left(d^4 + 6d^3 + 8d^2 + \sum_{P \in \mathcal{P}_N} \left(3{\rm Tr}[P^2]^2 + 6{\rm Tr}[P^4] \right)\right) \\ 
    &= \frac{1}{d^2} \left(d^4 + 6d^3 + 8d^2 + d^2(3d^2 + 6d)\right) \\
    &=  d^2 + 6d + 8 + 3d^2 + 6d \\
    &= 4d^2 + 12d + 8 \\
    &= 4(d+1)(d+2) \label{eq:QPiSym-calc}. 
\end{align}
Straightforward algebra using~\eqref{eq:alpha-beta-defs} and the fact that $M_{\rm lin}(|\psi\rangle) = 1 - d|\Xi(|\psi\rangle)|_2^2$ yields 
\begin{align}
    b^2 &= \frac{(d- 1 + M_{\rm lin}(|\psi\rangle))^2}{d^2(d^2 - 1)^2(d+2)^2(d+4)^2} \\
    ab &= \frac{(1-M_{\rm lin}(|\psi\rangle))(d - 1 + M_{\rm lin}(|\psi\rangle))}{4d^2(d^2 - 1)(d+1)(d+2)^2(d+4)} - 
    \frac{(d - 1 + M_{\rm lin}(|\psi\rangle))^2}{d^2(d^2 -1)^2(d+2)^2(d+4)^2} \\
    a^2 &= 
    \frac{(1 - M_{\rm lin}(|\psi\rangle))^2}{16d^2(d+1)^2(d+2)^2} -
    \frac{(1 - M_{\rm lin}(|\psi\rangle))(d - 1 + M_{\rm lin}(|\psi\rangle))}{2d^2(d^2 - 1)(d+1)(d+2)^2(d+4)} +
    \frac{(d- 1 + M_{\rm lin}(|\psi\rangle))^2}{d^2(d^2-1)^2(d+2)^2(d+4)^2}.
\end{align}
Substituting~\eqref{eq:QPiSym-calc}, the equation ${\rm Tr}[\Pi_4^{\rm sym}] = d(d+1)(d+2)(d+3)$, and the preceding expressions for $a^2, ab$, and $b^2$ into~\eqref{eq:projector-simplification}, we get that
\begin{align}
    (i) &= \frac{6(1 - M_{\rm lin}(|\psi\rangle))^2}{d^2(d+1)(d+2)} - \frac{48(1 - M_{\rm lin}(|\psi\rangle))(d-1 + M_{\rm lin}(|\psi\rangle))}{d^2(d^2 - 1)(d+2)(d+4)} +
    \frac{96(d-1+M_{\rm lin}(|\psi\rangle))^2}{d^2(d-1)^2(d+1)(d+2)(d+4)^2} \\ 
    &+ \frac{48(1-M_{\rm lin}(|\psi\rangle))(d-1+M_{\rm lin}(|\psi\rangle))}{d^2(d^2 - 1)(d+2)(d+4)} - \frac{192(d - 1 + M_{\rm lin}(|\psi\rangle))^2}{d^2(d-1)^2(d+1)(d+2)(d+4)^2} +
    \frac{24(d- 1 + M_{\rm lin}(|\psi\rangle))^2(d+3)}{d(d-1)^2(d+1)(d+2)(d+4)^2} \label{eq:i-calc} .
\end{align}

The evaluation of term $(ii)$ is simpler. We compute
\begin{align}
    (ii) &\coloneqq {\rm Tr}[\rho_C^{(4)}\rho_H^{(4)}] \\
    &= {\rm Tr}\left([aQ\Pi_4^{\rm sym} + b\Pi_4^{\rm sym}] \frac{\Pi_4^{\rm sym}}{{\rm Tr}[\Pi_4^{\rm sym}]}\right) \\
    &= 24\left(a \frac{{\rm Tr}[Q\Pi_4^{\rm sym}]}{{\rm Tr}[\Pi_4^{\rm sym}]} + b \frac{{\rm Tr}[\Pi_4^{\rm sym}]}{{\rm Tr}[\Pi_4^{\rm sym}]}\right) \\
    &= \frac{96a}{d(d+3)} + 24b \\
    &= \frac{24(1 - M_{\rm lin}(|\psi\rangle))}{d^2(d+1)(d+2)(d+3)} - \frac{96(d - 1 + M_{\rm lin}(|\psi\rangle))}{d^2(d^2 - 1)(d+2)(d+3)(d+4)} + \frac{24(d - 1 + M_{\rm lin}(|\psi\rangle))}{d(d^2 - 1)(d+2)(d+4)} \label{eq:ii-calc},
\end{align}
where we again rely on the fact that $\rho_C^{(4)} = a Q\Pi_4^{\rm sym} + b \Pi_4^{\rm sym}$, equation~\eqref{eq:QPiSym-calc}, and also $(\Pi_4^{\rm sym})^2 = 24\Pi_4^{\rm sym}$.

Finally, we note that term $(iii)$ simply evaluates to
\begin{equation}\label{eq:iii-calc}
    (iii) \coloneqq {\rm Tr}[(\rho_H^{(4)})^2] = \frac{24}{d(d+1)(d+2)(d+3)},
\end{equation}
as in our calculation for the projected ensemble. We could now combine equations~\eqref{eq:i-calc},~\eqref{eq:ii-calc}, and~\eqref{eq:iii-calc} to obtain a ten-term expression for $\|\rho_C^{(4)} - \rho_H^{(4)}\|_2^2 = (i) - 2(ii) + (iii)$. However, by simply considering the terms of leading order in $1/d$ for these expressions, it already becomes clear that
\begin{equation*}
    \left \|\langle C^{\otimes 4} \psi C^{\dagger \otimes 4} \rangle_{\mathcal{C}_N} - \rho_H^{(4)} \right \|_2^2 \sim \frac{6(1-M_{\rm lin}(|\psi\rangle))^2}{d^4}.
\end{equation*}

\end{document}